\documentclass[11pt,a4paper]{amsart}

\usepackage{amsmath,amsthm,amssymb}
\usepackage{epic,eepic}
\usepackage{epsfig,indentfirst}
\usepackage{float,array,varioref}

\restylefloat{figure}

\begin{document}
\title[Structure constants as determinants]
{${\mathcal{N}}\!=\!4$ SYM structure constants as determinants}
\author{Omar Foda}
\address{Department of Mathematics and Statistics,
         The University of Melbourne,
         Parkville, Victoria 3010, Australia}
\email{omar.foda@unimelb.edu.au}
\begin{abstract}
We obtain a determinant expression for the tree-level structure 
constant of three non-extremal single-trace operators in the 
$SU(2)$ sector of planar ${\mathcal{N}} \! = \! 4$ supersymmetric 
Yang-Mills theory.
\end{abstract}
\maketitle

\newcommand{\Y}{\mathcal{Y}}
\newcommand{\Tr}{{\rm Tr}}
\newcommand{\Det}{{\rm Det}}
\renewcommand{\d}{\partial}
 \newcommand{\<}{{\langle}}
\renewcommand{\>}{{\rangle}}
\newcommand{\cA}{{\cal A}}
\newcommand{\cB}{{\cal B}}
\newcommand{\cC}{{\cal C}}
\newcommand{\cH}{{\cal H}}
\newcommand{\cL}{{\cal L}}

\newcommand{\N}{\mathcal{N}}
\newcommand{\F}{\mathcal{F}}
\newcommand{\f}{\mathcal{f}}
\newcommand{\Q}{\mathcal Q}
\def\O{{\mathcal O}}
\def\cJ{{\cal J}}
\def\cN{{\cal N}}
\def\cF{{\cal F}}
\newcommand{\CC}{\field{C}}
\newcommand{\NN}{\field{N}}
\newcommand{\ZZ}{\field{Z}}

\newcommand{\B}{\mathcal B}
\newcommand{\D}{\mathcal D}
\newcommand{\M}{\mathcal M}
\newcommand{\R}{\mathcal R}
\newcommand{\T}{\mathcal T}
\newcommand{\U}{\mathcal U}
\newcommand{\Z}{\mathcal Z}

\renewcommand{\P}{{\mathcal P}}
\newtheorem{ca}{Figure}

\def\ll{ \left\lgroup}
\def\rr{\right\rgroup}

\newcommand{\0}{{\bf 0.}}
\newcommand{\1}{{\bf 1.}}
\newcommand{\2}{{\bf 2.}}
\newcommand{\3}{{\bf 3.}}
\newcommand{\4}{{\bf 4.}}
\newcommand{\5}{{\bf 5.}}
\newcommand{\6}{{\bf 6.}}
\newcommand{\7}{{\bf 7.}}
\newcommand{\8}{{\bf 8.}}
\newcommand{\9}{{\bf 9.}}

\def\nn{\nonumber}

\def\union{\mathop{\bigcup}}

\def\lprod{\mathop{\prod{\mkern-29.5mu}{\mathbf\longleftarrow}}}
\def\rprod{\mathop{\prod{\mkern-28.0mu}{\mathbf\longrightarrow}}}

\def\sul{\sum\limits}
\def\pl{\prod\limits}

\def\pd #1{\frac{\partial}{\partial #1}}
\def\const{{\rm const}}

\def\tr{\operatorname{tr}}
\def\Res{\operatorname{Res}}
\def\det{\operatorname{det}}

\newcommand{\infinity}{\infty}
\newcommand{\product }{\prod }

\hyphenation{boson-ic
             ferm-ion-ic
             para-ferm-ion-ic
             two-dim-ension-al
             two-dim-ension-al
             rep-resent-ative
             par-tition}

\renewcommand{\mod}{\textup{mod}\,}
\newcommand{\wt}{\text{wt}\,}

\hyphenation{And-rews
             config-ura-tion
             config-ura-tions
             Gor-don
             boson-ic
             ferm-ion-ic
             para-ferm-ion-ic
             two-dim-ension-al
             two-dim-ension-al}

\setcounter{section}{-1}

\section{Overview}
\label{section-overview}

This note is based on \cite{E1}, where a computationally tractable 
expression for a class of structure constants in $\N\! =\! 4$ 
supersymmetric Yang-Mills theory, SYM$_4$, is obtained, and on 
\cite{KMT,MW}, where a restricted version of Slavnov's scalar 
product in XXZ spin-$\frac{1}{2}$ chains, of which the XXX 
spin-$\frac{1}{2}$ chain discussed in this note is a special 
case, is discussed.

To put the result of this note in context, we start the rest of 
this section with a brief overview of some of the highlights of 
integrability in SYM$_4$, together with references to original 
works as well competent reviews.
Following that, we recall basic definitions from the theory of 
quantum integrable models that are needed to explain our result. 
We refer the reader to the literature for technical details. 
Finally, we outline our result and the contents of the rest 
of the sections.

\subsection{Integrability in planar SYM$_4$} 

The discovery of integrable structures, on both sides of
Maldacena's AdS/CFT correspondence \cite{maldacena}, is undoubtedly 
one of the major developments in mathematical physics in the past ten
years \cite{beisert-review}. This is not only because of the obvious 
intrinsic importance of building bridges between subjects that 
would otherwise remain unrelated, but also because integrability 
may be the right approach to put the correspondence on a rigorous 
footing. 

In this note, we restrict our attention to integrability in planar 
SYM$_4$ on the CFT side of AdS/CFT.  
The planar limit (the number of colours $N_c$ $\rightarrow$ 
$\infty$, the gauge coupling $g_{\textit{YM}}$ $\rightarrow$ 
$0$, while the \rq t Hooft coupling 
$\lambda = g^2_{\textit{YM}} N_c$ remains finite) allows 
SYM$_4$ to be integrable. It is possible that integrability 
persists beyond the planar limit, but at this stage, this is 
a wide open question.

\subsection{SYM$_4$ and spin chains. 1-loop results}
SYM$_4$ contains an $SO(6)$ invariant scalar sector, that consists 
of six real scalars $\phi_i$, $i \in \{1, \cdots, 6\}$.
In \cite{MZ}, Minahan and Zarembo showed that the action of 
the 1-loop dilatation operator $D$ on single-trace operators 
$\{\O\}$ with 1-loop conformal dimensions $\{\Delta_{\O}\}$, 
in the scalar sector map to the action of the Hamiltonian on 
states in an integrable periodic $SO(6)$ spin-chain with 
nearest-neighbour interactions. 

The single-trace operators $\{\O\}$ map to eigenstates of 
the spin-chain Hamiltonian. Their conformal dimensions 
$\{\Delta_{\O}\}$ map to the corresponding eigenvalues. 
In \cite{beisert0307}, Beisert extended the result of 
\cite{MZ} to all fundamental fields in SYM$_4$.

\subsection{SYM$_4$ and spin chains. Higher loop results}
The six scalar fields $\phi_i$, $i \in \{1, \cdots, 6\}$, 
can be combined into three charged scalars $\{X, Y, Z\}$ 
and their charge conjugates $\{\bar{X}, \bar{Y}, \bar{Z}\}$. 
Any two non-conjugate fields, such as $\{X, Y\}$, form 
a closed $SU(2)$ subsector.

In \cite{BKS}, Beisert, Kristjansen and Staudacher established 
integrability in the $SU(2)$ scalar sector, up to 3-loops.
However, beyond 1-loop order, the action of the dilatation 
operator on gauge-invariant states can no longer be 
represented in terms of a nearest-neighbour spin-chain 
Hamiltonian. 

In \cite{SS}, Serban and Staudacher matched the dilatation 
operator in the $SU(2)$ sector with higher Hamiltonians 
in the Inozemtsev model, which is a spin-chain with long 
range interactions, up to 3-loop level, and an asymptotic 
Bethe Ansatz was proposed to obtain the Bethe eigenstates 
and eigenvalues in the long chain-length limit, 
$L$ $\rightarrow$ $\infty$. These results can also be 
obtained using the Hubbard model 
\cite{rej-serban-staudacher, beisert0610}. 
But these models do not match the dilatation operator 
beyond 3-loop level and the final word on the integrable 
model that describes SYM$_4$ to all loop order remains 
to be written.

\subsection{All-sector, all-loop asymptotic Bethe 
Ansatz equations}

In \cite{BDS}, Beisert, Dippel and Staudacher proposed
asymptotic (valid with no corrections only in the long 
chain-length limit, $L \rightarrow \infty$) all-loop 
Bethe Ansatz equations in the $SU(2)$ sector. These 
equations require a {\it dressing factor} to match 
predictions made in the strong coupling limit.
In \cite{janik}, Janik proposed an equation that the 
dressing factor must satisfy. In \cite{BHL}, Beisert, 
Hernandez and Lopez solved Janik's equation.
In \cite{BES} Beisert, Eden and Staudacher showed 
that this solution has the right properties in the 
weak coupling limit. 
In \cite{BS0504}, Beisert and Staudacher proposed 
asymptotic Bethe 
Ansatz equations that hold for all sectors to all loops, 
in the $L \rightarrow \infty$ limit. This proposal was 
confirmed in \cite{beisert0511}. 

\subsection{Finite-size corrections}
The asymptotic Bethe Ansatz equations are 
valid without corrections only in the $L \rightarrow \infty$ 
limit. For long but finite length chains, we need to compute 
the finite size corrections. 

One approach to computing finite-size corrections is L\"{u}scher's 
method, introduced in the context of weak coupling integrability 
by Janik and Lukowsky \cite{JL} and applied by Bajnok and Janik 
\cite{BJ}. For an introduction to this method in AdS/CFT, 
see \cite{janik-review}. 
Another approach to finite-size corrections is the thermodynamic 
Bethe Ansatz, TBA, first considered in the AdS/CFT framework by 
Ambjorn, Janik and Kristjansen \cite{AJK}. It relies on the 
equivalence of a finite-size, zero-temperature theory to 
an infinite-size, finite-tempearture mirror theory. 
The ground state energy is then computed by solving 
sets of coupled nonlinear integral TBA equations 
\cite{bajnok-review}. 

TBA equations can be put in an elegant, universal form called 
Y-systems, which are systems of difference equations that appear 
in diverse topics in classical as well as quantum integrability. 
For a comprehensive review of Y-stsyems, see \cite{KNS}. 
For a review of applications of Y-systems in AdS/CFT, 
see \cite{GK}.

\subsection{Weakly-coupled, planar SYM$_4$. The $SU(2)$ 
scalar sector.} In this note, we restrict ourselves to 
weakly-coupled planar SYM$_4$, where perturbation theory 
in \lq t Hooft's coupling constant $\lambda$ is valid and 
we can consistently work up to 1-loop order. When this is 
the case, we can make use of mappings to integrable 
spin chains with nearest neighbour interactions, and 
conventional tools, such as the algebraic Bethe Ansatz 
apply. 

Furthermore, we deal only sectors with two complex scalars, 
so that the mapping is to $SU(2)$ spin-$\frac{1}{2}$ chains. 
It is only in the case of spin chains based on rank-1 Lie 
algebras that we have a determinant expression for the inner 
product of a Bethe eigenstate and a generic state \cite{slavnov}, 
which will be the main tool that we will use to obtain 
determinant expressions for structure constants.

\subsection{Conformal invariance and 2-point functions}
\label{conf-inv-2-pt}
Because SYM$_4$ is con\-form\-ally-invariant at the quantum 
level, it contains a basis of local gauge-invariant 
composite operators $\{\O\}$ such that each $\O_i$ 
$\in$ $\{\O\}$ is an eigenstate of the dilatation 
operator $D$, with a corresponding eigenvalue 
$\Delta_{\O_i}$, equal to the conformal dimension of $\O$. 
The 2-point function of $\O_i$ and $\O_j$ can be written 
as 
\begin{equation}
\<\O_i(x) \bar\O_j(y)\> 
=
\ll {\N_i \ \N_j} \rr^{1/2}
\frac{ \delta_{ij} } { |x - y|^{2 \Delta_i} }
\label{2-point}
\end{equation}
\noindent where 
$\bar\O_j$ is the Wick conjugate of $\O_i$, 
$\Delta_i$ is once again the conformal dimension 
of $\O_i$, and $\N_i$ is a normalization 
factor \footnote{Later, we will choose $\N_i$ 
to be (the square root of) the Gaudin norm of 
the corresponding spin-chain state.}.
The 2-point functions of $\{\O\}$ and their 
conformal dimensions $\{\Delta_{\O}\}$ are by 
now well-understood \cite{beisert-review}, 
and the next logical step is to study 3-point 
functions of $\{\O\}$ and their structure constants 
\cite{earlier-papers-on-3-pt-funs, E1, E2, E3}.
\subsection{3-point functions and structure constants} 
A 3-point function of basis local operators in SYN$_4$,  
is restricted by conformal symmetry to be of the form
\begin{multline}
\< \O_i (x_i) 
   \O_j (x_j) 
   \O_k (x_k) 
\> 
= 
\\
\ll
{\N_i \ \N_j \ \N_k}
\rr^{1/2}
\frac{
C_{ijk}
}
{
 |x_{ij}|^{\Delta_i + \Delta_j - \Delta_k} 
 |x_{jk}|^{\Delta_j + \Delta_k - \Delta_i} 
 |x_{ki}|^{\Delta_k + \Delta_i - \Delta_j}
}
\label{3-point}
\end{multline}
\noindent where $x_{ij} = x_i - x_j$, and $C_{ijk}$ is the structure 
constant. 

In this work, we restrict our attention to the weak-coupling limit 
where perturbation theory in the \lq t Hooft coupling constant 
$\lambda$ makes sense, and we can further restrict our analysis to 
1-loop level perturbation theory. In this limit, we can describe 
the integrability of SYM$_4$ in terms of spin-chains with 
nearest-neighbour interactions where conventional tools such as the 
algebraic Bethe Ansatz are most effective.

In \cite{E1}, Escobedo, Gromov, Sever and Vieira (EGSV) obtained 
an expression expressions for the structure constants of 
non-extremal single-trace operators in the scalar sector of SYM$_4$ 
that contains two charged scalars $\{Z, X\}$ and their conjugates 
$\{\bar{Z}, \bar{X}\}$. The EGSV expression is in terms of a sum 
over partitions of a set of rapidities into two distinct subsets.
In this paper, the sum expression of EGSV is evaluated in determinant 
form. This determinant turns out to be (a restriction of) the 
well-known Slavnov determinant in exact solutions in statistical 
mechanics. It is equal to the inner product of a Bethe eigenstate 
and a generic state in Heisenberg XXZ spin-$\frac{1}{2}$ chains.

\subsection{Rapidity variables, generic Bethe states and Bethe 
eigenstates} 
States in a closed length-$L$ XXX spin-$\frac{1}{2}$ 
chain\footnote{We 
restrict our attention to this spin chain, and use `spin chain' to refer 
to that.} depend on two sets of rapidity variables, auxiliary space 
rapidity variables, `auxiliary rapidities', and quantum space rapidity 
variables, `quantum rapidities'. When all quantum rapidities are set 
equal to the same constant value, the spin chain is `homogeneous'.
At each of the $L$ sites, there is a state variable, or equivalently, 
a spin variable, that is represented by an arrow that can be either up 
or down. A state with all spins up is a `reference state'\footnote{In 
this note, `state variables', `spin variables', and `arrows' can be 
used interchangeably.}. 
Initial and final generic Bethe states, $| \O \>$ and $\< \O |$ are 
created by the action of algebraic Bethe Ansatz (BA) operators 
on initial and final spin-chain reference states. 
They are characterized by auxiliary rapidities that are free 
variables, and they are not eigenstates of the spin-chain transfer 
matrix. Initial and final Bethe eigenstates, $| \O \>_{\beta}$ and 
${}_{\beta}\< \O |$, are also created by the action of BA operators 
on reference states. However, their auxiliary rapidities satisfy 
Bethe equations, and consequently, they are eigenstates of the 
spin-chain transfer 
matrix. We use the subscript $\beta$ to distinguish between these two 
types of states, and refer to them as `generic states' and `eigenstates', 
respectively.

\subsection{An expression for the structure constant} 
In \cite{E1}, Escobedo, Gromov, Sever and Vieira (EGSV) obtained 
a computationally tractable expression for the tree-level structure 
constant $c^{(0)}_{ijk}$ of three operators, $\O_i$, $i \in \{1, 2, 3\}$ 
with definite 1-loop anomalous conformal dimensions, in the $SU(2)$ 
sector in planar SYM$_4$. 
We restrict our attention to these operators 
in this sector of this theory, and use `operators' and 
`structure constants' in the sense of this restriction. 
We use $c^{(0)}_{123}$ rather than $c^{(0)}_{ijk}$ 
when this simplifies the notation with no loss of generality. 

In \cite{E1}, EGSV make use of the connection of weakly coupled 
SYM$_4$ \footnote{For an introduction to integrability 
in gauge and string theory, see \cite{beisert-review}.}
to integrable spin chains to map the operators $\O_i$, 
to eigenstates $| \O_i \>_{\beta}$.
Following that, they
{\bf 1.} Split each initial eigenstate  
$| \O_{i} \>_{\beta} $ into two initial generic states,  
$| \O_{i} \>_r$ and  
$| \O_{i} \>_l$, 
{\bf 2.} Map the three initial generic states 
$| \O_i \>_r$, to the corresponding 
three final generic states $_r\< \O_i |$, and finally
{\bf 3.} Compute the structure constants by taking scalar 
products of specific pairs of initial and final generic states. 

From the above outline one expects two complications. 
{\bf A.} From step {\bf 1}, one expects a sum over many possible 
ways of splitting each eigenstate into two generic states, and 
{\bf B.} From step {\bf 3}, one expects that are three non-trivial
scalar products to evaluate.
Both of these expectations turn out to be incorrect. 

\subsection{A constraint that leads to two simplifications} 
In formulating $c^{(0)}_{123}$ in BA terms, EGSV start with 
three initial eigenstates, 
$| \O_1 \>_{\beta}$,  
$| \O_2 \>_{\beta}$, and 
$| \O_3 \>_{\beta}$,  
characterized by sets of auxiliary rapidities 
$\{u\}_{\beta N_1}$, $\{v\}_{\beta N_2}$ and $\{w\}_{\beta N_3}$ 
with cardinalities 
$N_1$, $N_2$ and $N_3$, respectively, that satisfy Bethe 
equations\footnote{In this note, the set $\{u\}$ will always 
have cardinality $N_1$ and satisfy Bethe equations, hence the
subscript $\beta$. The sets $\{v\}$ and $\{w\}$ will have 
cardinalities $N_2$ and $N_3$. They satisfy Bethe equations, 
but this fact is not used, and their Bethe equations will play 
no role. 
The quantum rapidities $\{z\}$ will have cardinality $L$, and 
do not satisfy Bethe equations.}.  
The set $\{N\}$ $\equiv$ $\{N_1, N_2, N_3\}$ will appear 
frequently in the sequel.
Remarkably, it turns out that $N_1 = N_2 + N_3$. This constraint 
distinguishes the eigenstate $| \O_1 \>_{\beta}$, and admits 
{\it one and only one} way to split each eigenstate into left 
and right generic states. This removes complication {\bf A}. 
It also reduces the number of scalar products that one expects 
to evaluate. One scalar product is constrained to be between two 
reference states and therefore trivial. A second scalar product 
is constrained to be between two dual reference states (states 
with all spins down) and therefore straightforward to compute. 
Only one scalar product remains to be evaluated and this removes 
complication {\bf B}. 

\subsection{A generic scalar product that is a weighted sum} 
The remaining scalar product is generic in the sense that it 
involves two generic states with rapidities that do not satisfy 
Bethe equations, and neither is a reference or a dual reference
state. There is no simple expression (such as a determinant) for 
a generic scalar product, but using the commutation relations of 
the BA operators, one can express it as a manageable 
sum \cite{korepin-book}. 
EGSV use this sum form of the generic scalar product to obtain 
a computationally tractable weighted 
sum over all partitions of the set $\{u\}_{N_1}$ of cardinality 
$N_1$ into two sets $\alpha$ and $\bar{\alpha}$ of cardinality 
$N_2$ and $N_3$, respectively. 

\subsection{Bethe equations, Slavnov's scalar product and the result 
in this note} This note is based on the observation that 
$c^{(0)}_{123}$ as defined in \cite{E1} is (up to a factor) 
a restricted version of a Slavnov scalar product of a generic state 
and an eigenstate. This restricted version is discussed in \cite{KMT} 
and was used in \cite{MW} to obtain a recursive proof of the determinant 
expression of Slavnov's scalar product\footnote{In \cite{MW}, one can 
also find a representation of this restricted scalar product in terms 
of six-vertex model diagrams. We will use this representation in this 
note.}.

This observation allows us to implicitly use the Bethe equations 
satisfied by $\{u\}_{\beta N_1} $ to evaluate the EGSV weighted 
sum over partitions of 
$\{u\}_{\beta N_1}$, and to write $c^{(0)}_{123}$ as a determinant 
of an $(N_1 \! \times \! N_1)$-matrix\footnote{The auxiliary rapidities 
$\{v\}$ and $\{w\}$  also satisfy Bethe equations, but this fact 
is not used in this note.}. 

\subsection{Outline of contents} 

The subject of this note is at the intersection of supersymmetric 
Yang-Mills theory and integrable statistical mechanical models. 
We cannot review either of these topics in any technical detail.
Overall, we can only recall the very basics that are needed to 
obtain our result and refer the reader to \cite{E1,KMT,MW} for 
a more complete discussion and references to the original literature. 
On the other hand, our presentation is elementary. In particular, 
we rephrase the operator language of spin chains in terms of the 
diagrammatic language of the six vertex model, in the hope that 
this will make our arguments more accessible to readers with 
minimal background in quantum integrable models. 

In Section {\bf \ref{section-six-vertex}}, we review standard facts 
related to the rational six-vertex model, which is basically another 
way to consider XXX spin-$\frac{1}{2}$ chains, but as mentioned 
above, we find that the diagrams that represent the vertex model 
lattice configurations 
better suit our purposes. Following \cite{KMT,MW}, we introduce the 
$[L, N_1, N_2]$-configurations that will be central to our result.
In Section {\bf \ref{section-spin-chain}}, we review standard 
facts related to XXX spin-$\frac{1}{2}$ chains, and rephrase 
various ingredients of the BA solution in terms of vertex model 
lattice configurations. Following \cite{KMT,MW}, we introduce 
restricted versions $S[L, N_1, N_2]$ of Slavnov's scalar product, 
that can be evaluated in determinant form. 
$S[L, N_1, N_2]$ will turn out to be
the partition function of the $[L, N_1, N_2]$-configurations 
introduced in Section {\bf \ref{section-six-vertex}}.
In Section {\bf \ref{section-structure-constants}}, we recall 
the EGSV expression of the structure constants, and express it 
in terms of vertex model lattice configurations. 
In Section {\bf \ref{section-determinant-expression}}, we identify 
the weighted sum in the EGSV expression with the restricted Slavnov 
scalar product $S[L, N_1, N_2]$ introduced earlier, thereby showing 
that, 
up to a multiplicative factor, $c^{(0)}_{123}$ can be written as 
a determinant.
In Section {\bf \ref{section-comments}}, we collect a number 
of comments and remarks.

\section{The rational six-vertex model}
\label{section-six-vertex}

In this section, we recall the 2-dimensional rational six-vertex 
model in the absence of external fields. From now on, `six-vertex 
model' will refer to that.
It is equivalent to the XXX spin-$\frac{1}{2}$ chain that appears 
in \cite{E1}, but affords a diagrammatic representation that suits 
our purposes. We introduce quite a few terms to make this corresponds
clear and the presentation precise, but any reader with basic 
familiarity with quantum integrable models can skip all these.  

\subsection{Lattice lines, orientations, and rapidity variables}
\begin{figure}
\setlength{\unitlength}{0.0009cm}
\begin{picture}(5000,5000)(1500, 1500)
\thicklines
%
\path(2400,4800)(2400,1800)
\path(3000,4800)(3000,1800)
\path(3600,4800)(3600,1800)
\path(4200,4800)(4200,1800)
\path(4800,4800)(4800,1800)
%
\path(1800,4200)(5400,4200)
\path(1800,3600)(5400,3600)
\path(1800,3000)(5400,3000)
\path(1800,2400)(5400,2400)
%
%
\path(0600,4254)(1200,4254)
\whiten\path(840,4164)(1200,4254)(840,4344)(840,4164)
\path(0600,3654)(1200,3654)
\whiten\path(840,3564)(1200,3654)(840,3744)(840,3564)
\path(0600,3054)(1200,3054)
\whiten\path(840,2964)(1200,3054)(840,3144)(840,2964)
\path(0600,2454)(1200,2454)
\whiten\path(840,2364)(1200,2454)(840,2544)(840,2364)
%
%
\path(2400,5300)(2400,5900)
\whiten\path(2490,5610)(2400,5250)(2310,5610)(2490,5610)
\path(3000,5300)(3000,5900)
\whiten\path(3090,5610)(3000,5250)(2910,5610)(3090,5610)
\path(3600,5300)(3600,5900)
\whiten\path(3690,5610)(3600,5250)(3510,5610)(3690,5610)
\path(4200,5300)(4200,5900)
\whiten\path(4290,5610)(4200,5250)(4110,5610)(4290,5610)
\path(4800,5300)(4800,5900)
\whiten\path(4890,5610)(4800,5250)(4710,5610)(4890,5610)
\put(0000,4200){$u_1$}
\put(0000,2000){$u_{L_h}$}
\put(2300,6250){$z_1$}
%
\put(4700,6250){$z_{L_v}$}
\end{picture}
\caption{A square lattice with oriented lines and rapidity variables. 
Lattice lines are assigned the orientations indicated by the 
white arrows.}
\label{lattice}
\end{figure}
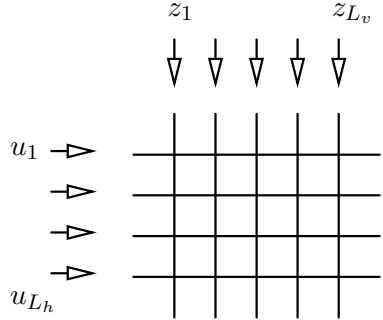

Consider a square lattice with $L_h$ horizontal lines and $L_v$ 
vertical lines that intersect at $L_h \! \times  \! L_v$ points.
There is no restriction, at this stage, on $L_h$ or $L_v$. 
We order the horizontal lines from top to bottom and assign the 
$i$-th line an orientation from left to right and a rapidity 
variable $u_i$. 
We order the vertical lines from left to right and assign
the $j$-th line an orientation from top to bottom and a rapidity 
variable $z_j$. See Figure {\bf \ref{lattice}}. The orientations 
that we assign to the lattice lines are matters of convention and 
are only meant to make the vertices of the six-vertex model, that 
we will introduce shortly, unambiguous. We orient the vertical 
lines from top to bottom to agree with the direction of the 
`spin set evolution' that we will introduce shortly.

%
\bigskip

\subsection{Bulk and boundary line segments, arrows, and vertices} 
Each lattice line is split into segments by all other lines that 
are perpendicular to it. Bulk segments that are attached to two 
intersection points, and boundary segments that are attached to 
one intersection point only.
Assign each segment an arrow that can point in either direction, 
and define the vertex $v_{ij}$ as a set of the three elements. 
\1 The intersection point of the $i$-th horizontal line and the 
$j$-th vertical line, 
\2 The four line segments attached to this intersection point, and 
\3 The arrows on these segments (regardless of their orientations). 
Assign $v_{ij}$ a weight that depends on the specific 
orientations of its arrows, and the rapidities $u_i$ and $z_j$
that flow through it.

\subsection{Six vertices that conserve `arrow flow'}
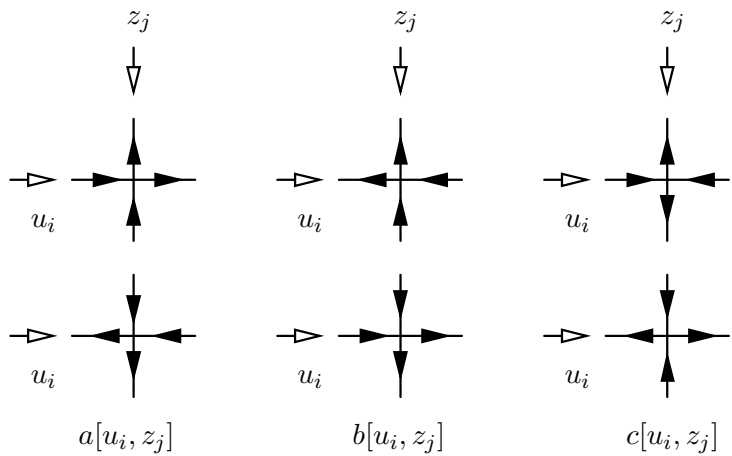
\begin{figure}
\setlength{\unitlength}{0.0009cm}
\begin{picture}(10000,7000)(1000, 1500)
\thicklines
%
\blacken\path( 1580,3490)( 1220,3400)( 1580,3310)( 1580,3490)
%
\blacken\path( 1220,5610)( 1580,5700)( 1220,5790)( 1220,5610)
%
\blacken\path( 2480,3490)( 2120,3400)( 2480,3310)( 2480,3490)
%
\blacken\path( 2120,5610)( 2480,5700)( 2120,5790)( 2120,5610)
%
\blacken\path( 5480,5790)( 5120,5700)( 5480,5610)( 5480,5790)
%
\blacken\path( 5120,3310)( 5480,3400)( 5120,3490)( 5120,3310)
%
\blacken\path( 6380,5790)( 6020,5700)( 6380,5610)( 6380,5790)
%
\blacken\path( 6020,3310)( 6380,3400)( 6020,3490)( 6020,3310)
%
\blacken\path( 9380,3490)( 9020,3400)( 9380,3310)( 9380,3490)
%
\blacken\path( 9020,5610)( 9380,5700)( 9020,5790)( 9020,5610)
%
\blacken\path( 9920,3310)(10280,3400)( 9920,3490)( 9920,3310)
%
\blacken\path(10280,5790)( 9920,5700)(10280,5610)(10280,5790)
%
\blacken\path( 1710,3160)( 1800,2800)( 1890,3160)( 1710,3160)
%
\blacken\path( 1710,3985)( 1800,3625)( 1890,3985)( 1710,3985)
%
\blacken\path( 5610,3160)( 5700,2800)( 5790,3160)( 5610,3160)
%
\blacken\path( 5610,4060)( 5700,3700)( 5790,4060)( 5610,4060)
%
\blacken\path( 9510,4060)( 9600,3700)( 9690,4060)( 9510,4060)
%
\blacken\path( 9510,5460)( 9600,5100)( 9690,5460)( 9510,5460)
%
\blacken\path( 1890,5040)( 1800,5400)( 1710,5040)( 1890,5040)
%
\blacken\path( 1890,5940)( 1800,6300)( 1720,5940)( 1890,5940)
%
\blacken\path( 5790,5040)( 5700,5400)( 5610,5040)( 5790,5040)
%
\blacken\path( 5790,5940)( 5700,6300)( 5620,5940)( 5790,5940)
%
\blacken\path( 9690,2740)( 9600,3100)( 9520,2740)( 9690,2740)
%
\blacken\path( 9690,5940)( 9600,6300)( 9510,5940)( 9690,5940)
%
\path(01800,7650)(01800,7025)
\path(01800,6600)(01800,4800)
\path(01800,4300)(01800,2500)
%
\path(05700,7650)(05700,7025)
\path(05700,6600)(05700,4800)
\path(05700,4300)(05700,2500)
%
\path(09600,7650)(09600,7025)
\path(09600,6600)(09600,4800)
\path(09600,4300)(09600,2500)
%
%
\path(00000,3400)(00600,3400)
\path(00000,5700)(00600,5700)
%
\path(00900,3400)(02700,3400)
\path(00900,5700)(02700,5700)
%
\path(03900,3400)(04500,3400)
\path(03900,5700)(04500,5700)
%
\path(04800,3400)(06600,3400)
\path(04800,5700)(06600,5700)
%
\path(07800,3400)(08400,3400)
\path(07800,5700)(08400,5700)
\path(08700,3400)(10500,3400)
\path(08700,5700)(10500,5700)
%
%
%
\whiten\path(1890,7360)(1800,7000)(1710,7360)(1890,7360)
\whiten\path(5790,7360)(5700,7000)(5610,7360)(5790,7360)
\whiten\path(9690,7360)(9600,7000)(9510,7360)(9690,7360)
%
%
\whiten\path(0260,3310)(0620,3400)(0260,3490)(0260,3310)
\whiten\path(0260,5610)(0620,5700)(0260,5790)(0260,5610)
\whiten\path(4160,3310)(4520,3400)(4160,3490)(4160,3310)
\whiten\path(4160,5610)(4520,5700)(4160,5790)(4160,5610)
\whiten\path(8060,3310)(8420,3400)(8060,3490)(8060,3310)
\whiten\path(8060,5610)(8420,5700)(8060,5790)(8060,5610)
\put(1000,1800){$a[u_i, z_j]$}
\put(5000,1800){$b[u_i, z_j]$}
\put(9000,1800){$c[u_i, z_j]$}
\put(0300,5000){$u_i$}
\put(4200,5000){$u_i$}
\put(8100,5000){$u_i$}
\put(0300,2700){$u_i$}
\put(4200,2700){$u_i$}
\put(8100,2700){$u_i$}
\put(1700,8000){$z_j$}
\put(5600,8000){$z_j$}
\put(9500,8000){$z_j$}
\end{picture}
\caption{The non-vanishing-weight vertices of the six-vertex model. 
Pairs of vertices in the same column share the weight that 
is shown below that column. 
The white arrows indicate the line orientations needed to 
specify the vertices without ambiguity.}
\label{six-vertices}
\end{figure}

Since every arrow can point in either direction, there are 
$2^4 = 16$ possible types of vertices. In this note, we are 
interested in a model such that only those vertices that 
conserves `arrow flow' (that is, the number of arrows that 
point toward the intersection point is equal to the number 
of arrows that point away from it) have non-zero weights. 
There are six such vertices. They are shown in 
Figure {\bf \ref{six-vertices}}. We assign these vertices 
non-vanishing weights. We assign the rest of the 16 possible 
vertices zero weights \cite{baxter-book}. 

In the rational six-vertex model, and in the absence of external 
fields, the six vertices with non-zero weights form three 
equal-weight pairs of vertices, as in Figure {\bf \ref{six-vertices}}. 
Two vertices that form a pair are related by reversing all arrows,
thus the vertex weights are invariant under reversing all arrows.
In the notation of Figure {\bf \ref{six-vertices}}, the weights 
of the rational six-vertex model, in the absence of external 
fields, are 
\begin{equation}
\label{weights}
a[u_i, z_j] =  \frac{(u_i - z_j + \eta)}
                    {(u_i - z_j       )}, 
\quad
b[u_i, z_j] =  1 
\quad
c[u_i, z_j] =  \frac{             \eta }
                    {(u_i - z_j       )}
\end{equation}

The assignment of weights in Equation {\bf \ref{weights}} 
satisfies unitarity, crossing symmetry, and most importantly 
the Yang-Baxter equations \cite{baxter-book}. It is not 
unique since one can multiply all weights by the same 
factor without changing the final physical results 
\footnote{The normalization of the vertex weights in 
Equation ({\bf \ref{weights}}) is different from that 
in \cite{KMT}. 
The latter is such that $b[u_a, z_j]$ $=$ $1$. 
We will comment on this again in Section 
{\bf \ref{section-structure-constants}}.}.

%
\bigskip

\subsection{Remarks} 
\1 The spin chain that is relevant to SYM$_4$ is 
homogeneous since all quantum rapidities are set equal to 
the same constant value $z$. In our conventions, $z$ $=$ 
$\frac{1}{2} \sqrt{-1}$. 
\2 The rational six-vertex model that corresponds to the 
homogeneous XXX spin-$\frac{1}{2}$ chain used in \cite{E1} 
will have, in our conventions, all vertical rapidity 
variables equal to $\frac{1}{2} \sqrt{-1}$. 
In this note, we start 
with inhomogeneous vertical rapidities, then take the 
homogeneous limit at the end.
\3 In a 2-dimensional vertex model with no external fields, 
the horizontal lines are on equal footing with the vertical 
lines. To make contact with spin chains, we will treat these 
two sets of differently.
\4 In figures in this note, a line segment with an arrow 
on it obviously indicates a definite arrow assignment. 
A line segment with no arrow on it implies a sum over 
both arrow assignments. 

\subsection{Weighted configurations and partition functions} 
By assigning every vertex $v_{ij}$ a weight $w_{ij}$, a vertex 
model lattice configuration with a definite assignment of arrows 
is assigned a weight equal to the product of the weights of its 
vertices. 
The partition function of a lattice configuration is the sum of 
the weights of all possible configurations that the vertices can 
take and that respect the boundary conditions. Since the vertex 
weights are invariant under reversal of all arrows, the partition 
functions is also invariant under reversal of all arrows.

\subsection{Rows of segments, spin systems, spin system states and 
net spin} 
`A row of segments' is a set of {\it vertical} line segments that 
start and/or end on the same horizontal line(s).
An $(L_v \! \times  \! L_h)$ six-vertex lattice configuration 
has $(L_v + 1)$ rows of segments. On every length-$L_h$ row of 
segments, one can assign a definite spin configuration, whereby 
each segment carries a spin variable (an arrow) that can point 
either up or down. 
A spin system on a specific row of segments is a set of all 
possible definite spin configurations that one can assign to 
that row. 
`A spin system state' is a one definite such configuration.
Two neighbouring spin systems (or spin system states) are 
separated by a horizontal lattice line. The spin systems on 
the top and the bottom rows of segments are initial and final 
spin systems, respectively. Consider a specific spin system 
state. Assign each up-spin the value $+1$ and each down-spin 
the value $-1$. The sum of these values is the net spin 
of this spin system state. In this note, we only consider 
six-vertex model configurations such that all elements in 
a spin system will the same net spin.

\subsection{Four types of horizontal lines} 
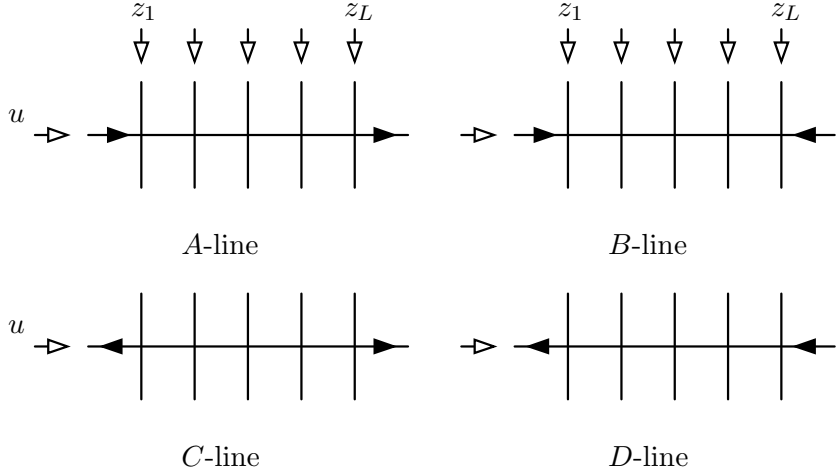
\begin{figure}
\setlength{\unitlength}{0.00035cm}
\begin{picture}(20000,18000)(2500,-13000)
\thicklines
\blacken\path(-1250,00250)(-1250,-0250)(-0530,00000)(-1250,00250)
\blacken\path(-0750,-7750)(-0750,-8250)(-1470,-8000)(-0750,-7750)
\blacken\path(08750,00250)(08750,-0250)(09470,00000)(08750,00250)
\blacken\path(08750,-7750)(08750,-8250)(09470,-8000)(08750,-7750)
\blacken\path(14750,00250)(14750,-0250)(15470,00000)(14750,00250)
\blacken\path(15250,-7750)(15250,-8250)(14530,-8000)(15250,-7750)
\blacken\path(25250,00250)(25250,-0250)(24530,00000)(25250,00250)
\blacken\path(25250,-7750)(25250,-8250)(24530,-8000)(25250,-7750)
\path(00000,-02000)(00000, 2000)
\path(00000, 03000)(00000, 4000)
\path(00000,-10000)(00000,-6000)
\path(02000,-02000)(02000, 2000)
\path(02000, 03000)(02000, 4000)
\path(02000,-10000)(02000,-6000)
\path(04000,-02000)(04000, 2000)
\path(04000, 03000)(04000, 4000)
\path(04000,-10000)(04000,-6000)
\path(06000,-02000)(06000, 2000)
\path(06000, 03000)(06000, 4000)
\path(06000,-10000)(06000,-6000)
\path(08000,-02000)(08000, 2000)
\path(08000, 03000)(08000, 4000)
\path(08000,-10000)(08000,-6000)
\path(12000, 00000)(13000, 0000)
\path(12000,-08000)(13000,-8000)
\path(14000, 00000)(26000, 0000)
\path(14000,-08000)(26000,-8000)
\path(16000,-02000)(16000, 2000)
\path(16000, 03000)(16000, 4000)
\path(16000,-10000)(16000,-6000)
\path(18000,-02000)(18000, 2000)
\path(18000, 03000)(18000, 4000)
\path(18000,-10000)(18000,-6000)
\path(20000,-02000)(20000, 2000)
\path(20000, 03000)(20000, 4000)
\path(20000,-10000)(20000,-6000)
\path(22000,-02000)(22000, 2000)
\path(22000, 03000)(22000, 4000)
\path(22000,-10000)(22000,-6000)
\path(24000,-02000)(24000, 2000)
\path(24000, 03000)(24000, 4000)
\path(24000,-10000)(24000,-6000)
\path(-2000, 00000)(10000, 0000)
\path(-2000,-08000)(10000,-8000)
\path(-4000, 00000)(-3000, 0000)
\path(-4000,-08000)(-3000,-8000)
\put(01500,-04500){$A$-line}
\put(17500,-04500){$B$-line}
\put(01500,-12500){$C$-line}
\put(17500,-12500){$D$-line}
\put(15600, 04500){$z_1$}
\put(23600, 04500){$z_L$}
\put(-0400, 04500){$z_1$}
\put(07600, 04500){$z_L$}
\put(-5000, 00500){$u$}
\put(-5000,-07500){$u$}
\whiten\path(01750,03500)(02250,03500)(02000,02780)(01750,03500)
\whiten\path(03750,03500)(04250,03500)(04000,02780)(03750,03500)
\whiten\path(05750,03500)(06250,03500)(06000,02780)(05750,03500)
\whiten\path(07750,03500)(08250,03500)(08000,02780)(07750,03500)
\whiten\path(12500,00250)(12500,-0250)(13220,00000)(12500,00250)
\whiten\path(12500,-7750)(12500,-8250)(13220,-8000)(12500,-7750)
\whiten\path(15750,03500)(16250,03500)(16000,02780)(15750,03500)
\whiten\path(17750,03500)(18250,03500)(18000,02780)(17750,03500)
\whiten\path(19750,03500)(20250,03500)(20000,02780)(19750,03500)
\whiten\path(21750,03500)(22250,03500)(22000,02780)(21750,03500)
\whiten\path(23750,03500)(24250,03500)(24000,02780)(23750,03500)
\whiten\path(-0250,03500)(00250,03500)(00000,02780)(-0250,03500)
\whiten\path(-3500,00250)(-3500,-0250)(-2780,00000)(-3500,00250)
\whiten\path(-3500,-7750)(-3500,-8250)(-2780,-8000)(-3500,-7750)
\end{picture}
\caption{There are four types of horizontal lines in a six-vertex model 
lattice configuration.}
\label{four-lines}
\end{figure}

Each horizontal line has two boundary segments. Each boundary 
segment has as an arrow that can point into the configuration 
or away from it. Accordingly, we can distinguish four types 
of horizontal lines, as in Figure {\bf \ref{four-lines}}. 
We will refer to them as $A$-, $B$-, $C$- and $D$-lines.

%
\bigskip

An important property of a horizontal line is how the net spin 
changes as one moves across it from top to bottom. Given that 
all vertices conserve 'arrow flow', one can easily show that, 
scanning a configuration from top to bottom, 
$B$-lines change the net spin by $-1$, 
$C$-lines increase it by $+1$, while 
$A$- and $D$-lines preserve the net spin. This can be easily 
understood by working out a few simple examples.

\subsection{Remarks} 
\1 There is of course no `time variable' in the six-vertex model, 
but one can think of a spin system as a dynamical system that 
evolves in discrete steps as one scans a lattice configuration 
from top to bottom. 
Starting from an initial spin set and scanning the configuration 
from top to bottom, one can think of the intermediate spin sets 
as consecutive states in the history of a dynamical system, 
ending with the final spin set. 
One can think of this evolution as caused by the action of 
the horizontal line elements. 

\2 In this note, all elements in a spin system, that live on a certain 
row of segments, have the same net spin. The reason is that vertically 
adjacent spin systems are separated by horizontal lines of a fixed 
type that change the net spin by the same amount ($\pm 1$) or keep 
it unchanged. Since we consider only lattice configurations with 
given horizontal lines (and do not sum over different types), the 
net spin of all elements in a spin system change by the same amount.

\subsection{Initial and final reference states, dual
reference states, and a variation}

An initial (respectively, final) reference state 
$| [L^{\wedge}] \>$ (respectively, $ \< [L^{\wedge}] | $)
is a spin system set on a top (respectively, bottom) row
of segments with $L$ arrows that are all up. 
An initial (final) dual reference state 
$| [L^{\vee}] \>$ ($ \< [L^{\vee}] | $)
is a spin system set on a top (bottom) row of segments with 
$L$ arrows that are all down. 
The state $ \< [ N_{3}^{\vee}, (L-N_3)^{\wedge}] |$ 
is a spin system state on a bottom row of segments with 
$L$ arrows such that the first $N_3$ 
arrows from the left are down, while the right $(L-N_3)$ 
arrows are up. We will not need the initial version 
of this state or their duals.

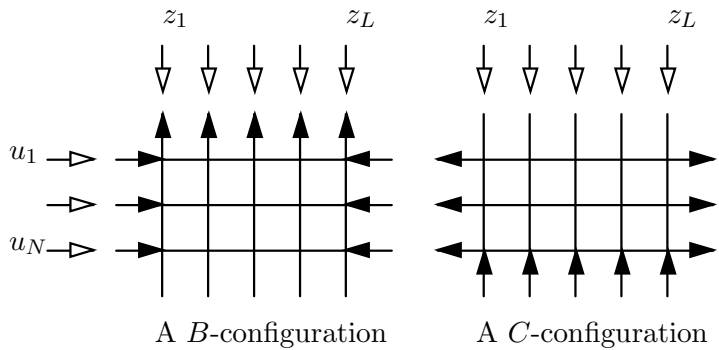
\begin{figure}
\setlength{\unitlength}{0.0010cm}
\begin{picture}(10000,4500)(1000,-0750)
\thicklines
\put(2100,3600){$z_1$}
\put(4500,3600){$z_L$}
\put(6300,3600){$z_1$}
\put(8700,3600){$z_L$}
\put(2000,-0600){A $B$-configuration}
\put(6200,-0600){A $C$-configuration}
\put(100,0610){$u_N$}
\put(100,1810){$u_1$}
\path(0600,0600)(1200,0600)
\path(0600,1200)(1200,1200)
\path(0600,1800)(1200,1800)
\path(1500,0600)(5100,0600)
\path(1500,1200)(5100,1200)
\path(1500,1800)(5100,1800)
%
%
\path(2100,2400)(2100,0000)
\path(2100,3300)(2100,3000)
\path(2700,2400)(2700,0000)
\path(2700,3300)(2700,3000)
\path(3300,2400)(3300,0000)
\path(3300,3300)(3300,3000)
\path(3900,2400)(3900,0000)
\path(3900,3300)(3900,3000)
\path(4500,2400)(4500,0000)
\path(4500,3300)(4500,3000)
\path(5700,0600)(9300,0600)
\path(5700,1200)(9300,1200)
\path(5700,1800)(9300,1800)
\path(6300,2400)(6300,0000)
\path(6300,3300)(6300,3000)
\path(6900,2400)(6900,0000)
\path(6900,3300)(6900,3000)
\path(7500,2400)(7500,0000)
\path(7500,3300)(7500,3000)
\path(8100,2400)(8100,0000)
\path(8100,3300)(8100,3000)
\path(8700,2400)(8700,0000)
\path(8700,3300)(8700,3000)
\blacken\path(1800,0700)(2100,0610)(1800,0520)(1800,0700)
\blacken\path(1800,1300)(2100,1210)(1800,1120)(1800,1300)
\blacken\path(1800,1900)(2100,1810)(1800,1720)(1800,1900)
\blacken\path(2020,2100)(2110,2400)(2200,2100)(2020,2100)
\blacken\path(2620,2100)(2710,2400)(2800,2100)(2620,2100)
\blacken\path(3220,2100)(3310,2400)(3400,2100)(3220,2100)
\blacken\path(3820,2100)(3910,2400)(4000,2100)(3820,2100)
\blacken\path(4420,2100)(4510,2400)(4600,2100)(4420,2100)
\blacken\path(4800,0520)(4500,0610)(4800,0700)(4800,0520)
\blacken\path(4800,1120)(4500,1210)(4800,1300)(4800,1120)
\blacken\path(4800,1720)(4500,1810)(4800,1900)(4800,1720)
\blacken\path(6000,0700)(5700,0610)(6000,0520)(6000,0700)
\blacken\path(6000,1300)(5700,1210)(6000,1120)(6000,1300)
\blacken\path(6000,1900)(5700,1810)(6000,1720)(6000,1900)
\blacken\path(6220,0300)(6310,0600)(6400,0300)(6220,0300)
\blacken\path(6820,0300)(6910,0600)(7000,0300)(6820,0300)
\blacken\path(7420,0300)(7510,0600)(7600,0300)(7420,0300)
\blacken\path(8020,0300)(8110,0600)(8200,0300)(8020,0300)
\blacken\path(8620,0300)(8710,0600)(8800,0300)(8620,0300)
\blacken\path(9000,0520)(9300,0610)(9000,0700)(9000,0520)
\blacken\path(9000,1120)(9300,1210)(9000,1300)(9000,1120)
\blacken\path(9000,1720)(9300,1810)(9000,1900)(9000,1720)
\whiten\path(0900,0520)(1200,0610)(0900,0700)(0900,0520)
\whiten\path(0900,1120)(1200,1210)(0900,1300)(0900,1120)
\whiten\path(0900,1720)(1200,1810)(0900,1900)(0900,1720)
\whiten\path(2020,3000)(2110,2700)(2200,3000)(2020,3000)
\whiten\path(2620,3000)(2710,2700)(2800,3000)(2620,3000)
\whiten\path(3220,3000)(3310,2700)(3400,3000)(3220,3000)
\whiten\path(3820,3000)(3910,2700)(4000,3000)(3820,3000)
\whiten\path(4420,3000)(4510,2700)(4600,3000)(4420,3000)
\whiten\path(6220,3000)(6310,2700)(6400,3000)(6220,3000)
\whiten\path(6820,3000)(6910,2700)(7000,3000)(6820,3000)
\whiten\path(7420,3000)(7510,2700)(7600,3000)(7420,3000)
\whiten\path(8020,3000)(8110,2700)(8200,3000)(8020,3000)
\whiten\path(8620,3000)(8710,2700)(8800,3000)(8620,3000)
\end{picture}
%
\caption{On the left, a $B$-configuration, generated by the action 
of $N$ $B$-lines on an initial length-$L$ reference state, 
$N \leq L$. 
A weighted sum over all possible configurations of segments 
with no arrows is implied. On the right, the corresponding 
$C$-configuration.}
%
\label{initial-final-state}
\end{figure}

\subsection{Four types of configurations} 

\1 A $B$-configuration is a lattice configuration with 
$L$ vertical lines and $N$ horizontal lines, 
$N \leq L$, such that
{\bf A.} The initial spin system is an initial reference state
$| [L^{\wedge}] \>$, and
{\bf B.} All horizontal lines $B$-lines. 
An example is on the left hand side of Figure 
{\bf \ref{initial-final-state}}.

\2 A $C$-configuration is a lattice configuration with 
$L$ vertical lines and $N$ horizontal lines, 
$N \leq L$, such that
{\bf A.} All horizontal lines are $C$-lines, and 
{\bf B.} The final spin system is a final reference state
$\< [L^{\wedge}] |$.
An example is on the right hand side of Figure 
{\bf \ref{initial-final-state}}.

%
\bigskip

\3 A $BC$-configuration is a lattice configuration with 
$L$ vertical lines and $2N_1$ horizontal lines, $0 \leq N_1 \leq L$,
such that
{\bf A.} The initial spin system is an initial reference state
$| [ L^{\wedge} ] \>$, 
{\bf B.} The first $N_1$ horizontal lines from top to bottom are $B$-lines,
{\bf C.} The following $N_1$ horizontal lines are $C$-lines,
{\bf D.} The final spin system is a final reference state
$ \< [ L^{\wedge} ] |$. See Figure {\bf \ref{bc-0}}\footnote{For 
visual clarity, we have allowed for a gap between 
the $B$-lines and the $C$-lines in Figure {\bf \ref{bc-0}}. 
There is also a gap between the $N_3$-th and $(N_3 +1)$-st vertical 
lines, where $N_3 = 3$ in the example shown, that indicates 
separate portions of the lattice that will be relevant shortly. 
The reader should ignore this at this stage.}.

\4 An $[L, N_1, N_2]$-configuration, $0 \leq N_2 \leq N_1$, is 
identical to a $BC$-configuration except that it has $N_1$ 
$B$-lines, and $N_2$ $C$-lines.  
When $N_3 = N_1 - N_2 = 0$, we evidently recover a $BC$-configuration.  
The case $N_2 = 0$ will be discussed below. 
For intermediate values of $N_2$, we obtain 
restricted $BC$-configurations whose partition functions
will turn out to be essentially the structure constants.

\subsection{$[L, N_1, N_2]$-configurations as restrictions 
of $BC$-configurations}
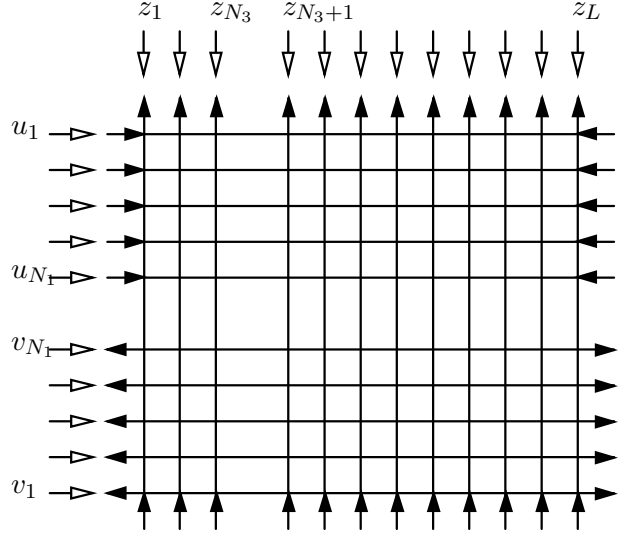
\begin{figure}
\setlength{\unitlength}{0.0008cm}
\begin{picture}(10000,09000)(0000,0750)
\thicklines
\path(0600,1500)(9000,1500)
\path(0600,2100)(9000,2100)
\path(0600,2700)(9000,2700)
\path(0600,3300)(9000,3300)
\path(0600,3900)(9000,3900)
\path(0600,5100)(9000,5100)
\path(0600,5700)(9000,5700)
\path(0600,6300)(9000,6300)
\path(0600,6900)(9000,6900)
\path(0600,7500)(9000,7500)
\path(1210,8100)(1210,0900)
\path(1800,8100)(1800,0900)
\path(2400,8100)(2400,0900)
\path(3600,8100)(3600,0900)
\path(4200,8100)(4200,0900)
\path(4800,8100)(4800,0900)
\path(5400,8100)(5400,0900)
\path(6000,8100)(6000,0900)
\path(6600,8100)(6600,0900)
\path(7200,8100)(7200,0900)
\path(7800,8100)(7800,0900)
\path(8400,8100)(8400,0900)
\blacken\path(0900,1590)(0600,1500)(0900,1410)(0900,1590)
\blacken\path(0900,2190)(0600,2100)(0900,2010)(0900,2190)
\blacken\path(0900,2790)(0600,2700)(0900,2610)(0900,2790)
\blacken\path(0900,3390)(0600,3300)(0900,3210)(0900,3390)
\blacken\path(0900,3990)(0600,3900)(0900,3810)(0900,3990)
\blacken\path(0900,5010)(1200,5110)(0900,5190)(0900,5010)
\blacken\path(0900,5610)(1200,5700)(0900,5790)(0900,5610)
\blacken\path(0900,6210)(1210,6300)(0900,6390)(0900,6210)
\blacken\path(0900,6810)(1200,6900)(0900,6990)(0900,6810)
\blacken\path(0900,7410)(1200,7500)(0900,7590)(0900,7410)
\blacken\path(1300,1200)(1210,1500)(1120,1200)(1300,1200)
\blacken\path(1300,7800)(1210,8100)(1120,7800)(1300,7800)
\blacken\path(1900,1200)(1810,1500)(1720,1200)(1900,1200)
\blacken\path(1900,7800)(1810,8100)(1720,7800)(1900,7800)
\blacken\path(2500,1200)(2410,1500)(2320,1200)(2500,1200)
\blacken\path(2500,7800)(2410,8100)(2320,7800)(2500,7800)
\blacken\path(3700,1200)(3610,1500)(3520,1200)(3700,1200)
\blacken\path(3700,7800)(3610,8100)(3520,7800)(3700,7800)
\blacken\path(4300,1200)(4210,1500)(4120,1200)(4300,1200)
\blacken\path(4300,7800)(4210,8100)(4120,7800)(4300,7800)
\blacken\path(4900,1200)(4810,1506)(4720,1200)(4900,1200)
\blacken\path(4900,7800)(4810,8100)(4720,7800)(4900,7800)
\blacken\path(5500,1200)(5410,1500)(5320,1200)(5500,1200)
\blacken\path(5500,7800)(5410,8100)(5320,7800)(5500,7800)
\blacken\path(6100,1200)(6010,1500)(5920,1200)(6100,1200)
\blacken\path(6100,7800)(6010,8100)(5920,7800)(6100,7800)
\blacken\path(6700,1200)(6610,1500)(6520,1200)(6700,1200)
\blacken\path(6700,7800)(6610,8100)(6520,7800)(6700,7800)
\blacken\path(7300,1200)(7210,1500)(7120,1200)(7300,1200)
\blacken\path(7300,7800)(7210,8100)(7120,7800)(7300,7800)
\blacken\path(7900,1200)(7810,1500)(7720,1200)(7900,1200)
\blacken\path(7900,7800)(7810,8100)(7720,7800)(7900,7800)
\blacken\path(8500,1200)(8410,1500)(8320,1200)(8500,1200)
\blacken\path(8500,7800)(8410,8100)(8320,7800)(8500,7800)
\blacken\path(8700,1400)(9000,1490)(8700,1580)(8700,1400)
\blacken\path(8700,2000)(9000,2090)(8700,2180)(8700,2000)
\blacken\path(8700,2600)(9000,2690)(8700,2780)(8700,2600)
\blacken\path(8700,3200)(9000,3290)(8700,3380)(8700,3200)
\blacken\path(8700,3800)(9000,3890)(8700,3980)(8700,3800)
\blacken\path(8700,5200)(8400,5110)(8700,5020)(8700,5200)
\blacken\path(8700,5800)(8400,5710)(8700,5620)(8700,5800)
\blacken\path(8700,6400)(8400,6310)(8700,6220)(8700,6400)
\blacken\path(8700,7000)(8400,6910)(8700,6820)(8700,7000)
\blacken\path(8700,7600)(8400,7510)(8700,7420)(8700,7600)
%
%
\path(-0360,1500)(0000,1500)
\path(-0360,2100)(0000,2100)
\path(-0360,2700)(0000,2700)
\path(-0360,3300)(0000,3300)
\path(-0360,3900)(0000,3900)
\put(-1000,1500){$v_1$}
\put(-1000,3900){$v_{N_1}$}
\whiten\path(0000,1410)(0360,1500)(0000,1590)(0000,1410)
\whiten\path(0000,2010)(0360,2100)(0000,2190)(0000,2010)
\whiten\path(0000,2610)(0360,2700)(0000,2790)(0000,2610)
\whiten\path(0000,3210)(0360,3300)(0000,3390)(0000,3210)
\whiten\path(0000,3810)(0360,3900)(0000,3990)(0000,3810)
\path(-0360,5100)(0000,5100)
\path(-0360,5700)(0000,5700)
\path(-0360,6300)(0000,6300)
\path(-0360,6900)(0000,6900)
\path(-0360,7500)(0000,7500)
\put(-1000,5100){$u_{N_1}$}
\put(-1000,7500){$u_1$}
\whiten\path(0000,5010)(0360,5100)(0000,5190)(0000,5010)
\whiten\path(0000,5610)(0360,5700)(0000,5790)(0000,5610)
\whiten\path(0000,6210)(0360,6300)(0000,6390)(0000,6210)
\whiten\path(0000,6810)(0360,6900)(0000,6990)(0000,6810)
\whiten\path(0000,7410)(0360,7500)(0000,7590)(0000,7410)
%
%
\path(1200,9220)(1200,8500)
\path(1800,9220)(1800,8500)
\path(2400,9220)(2400,8500)
\path(3600,9220)(3600,8500)
\path(4200,9220)(4200,8500)
\path(4800,9220)(4800,8500)
\path(5400,9220)(5400,8500)
\path(6000,9220)(6000,8500)
\path(6600,9220)(6600,8500)
\path(7200,9220)(7200,8500)
\path(7800,9220)(7800,8500)
\path(8400,9220)(8400,8500)
\put(1100,9500){$z_{1  } $}
\put(2300,9500){$z_{N_3  } $}
\put(3500,9500){$z_{N_3 +1} $}
\put(8300,9500){$z_{L  } $}
\whiten\path(1290,8860)(1200,8500)(1110,8860)(1290,8860)
\whiten\path(1890,8860)(1800,8500)(1710,8860)(1890,8860)
\whiten\path(2490,8860)(2400,8500)(2310,8860)(2490,8860)
\whiten\path(3690,8860)(3600,8500)(3510,8860)(3690,8860)
\whiten\path(4290,8860)(4200,8500)(4110,8860)(4290,8860)
\whiten\path(4890,8860)(4800,8500)(4710,8860)(4890,8860)
\whiten\path(5490,8860)(5400,8500)(5310,8860)(5490,8860)
\whiten\path(6090,8860)(6000,8500)(5910,8860)(6090,8860)
\whiten\path(6690,8860)(6600,8500)(6510,8860)(6690,8860)
\whiten\path(7290,8860)(7200,8500)(7110,8860)(7290,8860)
\whiten\path(7890,8860)(7800,8500)(7710,8860)(7890,8860)
\whiten\path(8490,8860)(8400,8500)(8310,8860)(8490,8860)
\end{picture}
%
\caption{A six-vertex model $BC$-configuration. 
$L   \!  = \! 12$, and 
$N_1   \!  = \!  5$, or equivalently 
$L_h \!  = \!  2  \! \times  \! 5 =\! 10$ and 
$L_v \! =  \! 12$.
The top    $N$ horizontal lines represent $B$-operators. 
The bottom $N$ horizontal lines represent $C$-operators.
The initial (top) as well as the final (bottom) boundary 
spin systems are reference states.} 
%
%
\label{bc-0}
\end{figure}

Consider a $BC$-configuration with no restrictions.  
To be specific, let us consider the configuration in 
Figure {\bf \ref{bc-0}}, where $N_1 = 5$ and $L = 12$. 
Consider the vertex at the bottom-left corner. 
For convenience, we label the $\{v\}$ rapidities from 
bottom to top. The $\{u\}$ rapidities are labeled from 
top to bottom as before. 

From Figure {\bf \ref{six-vertices}}, it is easy to see that
this can be either a $b$- or a $c$-vertex. Since the $\{v\}$ 
variables are free, set $v_1 = z_1$, thereby setting the 
weight of all configurations with a $b$-vertex at this 
corner to zero, and forcing the vertex at this corner to be 
$c$-vertex.

Referring to Figure {\bf \ref{six-vertices}} again, one 
can see that not only is the corner vertex forced to be 
type-$c$, but the orientations of all arrows on 
the horizontal lattice line with rapidity $v_1$, as well 
all all arrows on the vertical line with rapidity $z_1$ 
but below the horizontal line with rapidity $u_1$ are 
also frozen to fixed values.

%
\bigskip

\begin{figure}
\thicklines
\setlength{\unitlength}{0.0008cm}
\begin{picture}(10000,09000)(0000,0750)
\path(0600,1500)(9000,1500)
\path(0600,2100)(9000,2100)
\path(0600,2700)(9000,2700)
\path(0600,3300)(9000,3300)
\path(0600,3900)(9000,3900)
\path(0600,5100)(9000,5100)
\path(0600,5700)(9000,5700)
\path(0600,6300)(9000,6300)
\path(0600,6900)(9000,6900)
\path(0600,7500)(9000,7500)
\path(1200,8100)(1200,0900)
\path(1800,8100)(1800,0900)
\path(2400,8100)(2400,0900)
\path(3600,8100)(3600,0900)
\path(4200,8100)(4200,0900)
\path(4800,8100)(4800,0900)
\path(5400,8100)(5400,0900)
\path(6000,8100)(6000,0900)
\path(6600,8100)(6600,0900)
\path(7200,8100)(7200,0900)
\path(7800,8100)(7800,0900)
\path(8400,8100)(8400,0900)
\blacken\path(0900,1590)(0600,1500)(0900,1410)(0900,1590)
\blacken\path(0900,2190)(0600,2100)(0900,2010)(0900,2190)
\blacken\path(0900,2790)(0600,2700)(0900,2610)(0900,2790)
\blacken\path(0900,3390)(0600,3300)(0900,3210)(0900,3390)
\blacken\path(0900,3990)(0600,3900)(0900,3810)(0900,3990)
\blacken\path(0900,5010)(1200,5110)(0900,5190)(0900,5010)
\blacken\path(0900,5610)(1200,5700)(0900,5790)(0900,5610)
\blacken\path(0900,6210)(1210,6300)(0900,6390)(0900,6210)
\blacken\path(0900,6810)(1200,6900)(0900,6990)(0900,6810)
\blacken\path(0900,7410)(1200,7500)(0900,7590)(0900,7410)
\blacken\path(1300,1200)(1210,1500)(1120,1200)(1300,1200)
\blacken\path(1300,7800)(1210,8100)(1120,7800)(1300,7800)
\blacken\path(1900,1200)(1810,1500)(1720,1200)(1900,1200)
\blacken\path(1900,7800)(1810,8100)(1720,7800)(1900,7800)
\blacken\path(2500,1200)(2410,1500)(2320,1200)(2500,1200)
\blacken\path(2500,7800)(2410,8100)(2320,7800)(2500,7800)
\blacken\path(3700,1200)(3610,1500)(3520,1200)(3700,1200)
\blacken\path(3700,7800)(3610,8100)(3520,7800)(3700,7800)
\blacken\path(4300,1200)(4210,1500)(4120,1200)(4300,1200)
\blacken\path(4300,7800)(4210,8100)(4120,7800)(4300,7800)
\blacken\path(4900,1200)(4810,1506)(4720,1200)(4900,1200)
\blacken\path(4900,7800)(4810,8100)(4720,7800)(4900,7800)
\blacken\path(5500,1200)(5410,1500)(5320,1200)(5500,1200)
\blacken\path(5500,7800)(5410,8100)(5320,7800)(5500,7800)
\blacken\path(6100,1200)(6010,1500)(5920,1200)(6100,1200)
\blacken\path(6100,7800)(6010,8100)(5920,7800)(6100,7800)
\blacken\path(6700,1200)(6610,1500)(6520,1200)(6700,1200)
\blacken\path(6700,7800)(6610,8100)(6520,7800)(6700,7800)
\blacken\path(7300,1200)(7210,1500)(7120,1200)(7300,1200)
\blacken\path(7300,7800)(7210,8100)(7120,7800)(7300,7800)
\blacken\path(7900,1200)(7810,1500)(7720,1200)(7900,1200)
\blacken\path(7900,7800)(7810,8100)(7720,7800)(7900,7800)
\blacken\path(8500,1200)(8410,1500)(8320,1200)(8500,1200)
\blacken\path(8500,7800)(8410,8100)(8320,7800)(8500,7800)
\blacken\path(8700,1400)(9000,1490)(8700,1580)(8700,1400)
\blacken\path(8700,2000)(9000,2090)(8700,2180)(8700,2000)
\blacken\path(8700,2600)(9000,2690)(8700,2780)(8700,2600)
\blacken\path(8700,3200)(9000,3290)(8700,3380)(8700,3200)
\blacken\path(8700,3800)(9000,3890)(8700,3980)(8700,3800)
\blacken\path(8700,5200)(8400,5110)(8700,5020)(8700,5200)
\blacken\path(8700,5800)(8400,5710)(8700,5620)(8700,5800)
\blacken\path(8700,6400)(8400,6310)(8700,6220)(8700,6400)
\blacken\path(8700,7000)(8400,6910)(8700,6820)(8700,7000)
\blacken\path(8700,7600)(8400,7510)(8700,7420)(8700,7600)
%
%
\path(-0360,1500)(0000,1500)
\path(-0360,2100)(0000,2100)
\path(-0360,2700)(0000,2700)
\path(-0360,3300)(0000,3300)
\path(-0360,3900)(0000,3900)
\put(-1000,3900){$v_{ N_1           }$}
\put(-1000,2700){$v_{ N_3           }$}
\put(-1000,1500){$v_{             1 }$}
\whiten\path(0000,1410)(0360,1500)(0000,1590)(0000,1410)
\whiten\path(0000,2010)(0360,2100)(0000,2190)(0000,2010)
\whiten\path(0000,2610)(0360,2700)(0000,2790)(0000,2610)
\whiten\path(0000,3210)(0360,3300)(0000,3390)(0000,3210)
\whiten\path(0000,3810)(0360,3900)(0000,3990)(0000,3810)
\path(-0360,5100)(0000,5100)
\path(-0360,5700)(0000,5700)
\path(-0360,6300)(0000,6300)
\path(-0360,6900)(0000,6900)
\path(-0360,7500)(0000,7500)
\put(-1000,5100){$u_{N_1}$}
\put(-1000,7500){$u_{  1}$}
\whiten\path(0000,5010)(0360,5100)(0000,5190)(0000,5010)
\whiten\path(0000,5610)(0360,5700)(0000,5790)(0000,5610)
\whiten\path(0000,6210)(0360,6300)(0000,6390)(0000,6210)
\whiten\path(0000,6810)(0360,6900)(0000,6990)(0000,6810)
\whiten\path(0000,7410)(0360,7500)(0000,7590)(0000,7410)
%
%
\path(1200,9220)(1200,8500)
\path(1800,9220)(1800,8500)
\path(2400,9220)(2400,8500)
\path(3600,9220)(3600,8500)
\path(4200,9220)(4200,8500)
\path(4800,9220)(4800,8500)
\path(5400,9220)(5400,8500)
\path(6000,9220)(6000,8500)
\path(6600,9220)(6600,8500)
\path(7200,9220)(7200,8500)
\path(7800,9220)(7800,8500)
\path(8400,9220)(8400,8500)
\put(1100,9500){$z_1   $}
\put(2300,9500){$z_{ N_3    }$}
\put(3500,9500){$z_{ N_3 + 1 }$}
\put(8300,9500){$z_{L}$}
\whiten\path(1290,8860)(1200,8500)(1110,8860)(1290,8860)
\whiten\path(1890,8860)(1800,8500)(1710,8860)(1890,8860)
\whiten\path(2490,8860)(2400,8500)(2310,8860)(2490,8860)
\whiten\path(3690,8860)(3600,8500)(3510,8860)(3690,8860)
\whiten\path(4290,8860)(4200,8500)(4110,8860)(4290,8860)
\whiten\path(4890,8860)(4800,8500)(4710,8860)(4890,8860)
\whiten\path(5490,8860)(5400,8500)(5310,8860)(5490,8860)
\whiten\path(6090,8860)(6000,8500)(5910,8860)(6090,8860)
\whiten\path(6690,8860)(6600,8500)(6510,8860)(6690,8860)
\whiten\path(7290,8860)(7200,8500)(7110,8860)(7290,8860)
\whiten\path(7890,8860)(7800,8500)(7710,8860)(7890,8860)
\whiten\path(8490,8860)(8400,8500)(8310,8860)(8490,8860)
%
%
\blacken\path(1320,1410)(1680,1500)(1320,1590)(1320,1410)
\blacken\path(1920,1410)(2280,1500)(1920,1590)(1920,1410)
\blacken\path(2820,1410)(3180,1500)(2820,1590)(2820,1410)
\blacken\path(3720,1410)(4080,1500)(3720,1590)(3720,1410)
\blacken\path(4320,1410)(4680,1500)(4320,1590)(4320,1410)
\blacken\path(4920,1410)(5280,1500)(4920,1590)(4920,1410)
\blacken\path(5520,1410)(5880,1500)(5520,1590)(5520,1410)
\blacken\path(6120,1410)(6480,1500)(6120,1590)(6120,1410)
\blacken\path(6720,1410)(7080,1500)(6720,1590)(6720,1410)
\blacken\path(7320,1410)(7680,1500)(7320,1590)(7320,1410)
\blacken\path(7920,1410)(8280,1500)(7920,1590)(7920,1410)
%
%
\blacken\path(1680,2010)(1320,2100)(1680,2190)(1680,2010)
\blacken\path(1680,2610)(1320,2700)(1680,2790)(1680,2610)
\blacken\path(1680,3210)(1320,3300)(1680,3390)(1680,3210)
\blacken\path(1680,3810)(1320,3900)(1680,3990)(1680,3810)
%
%
\blacken\path(1920,2010)(2280,2100)(1920,2190)(1920,2010)
\blacken\path(2280,2610)(1920,2700)(2280,2790)(2280,2610)
\blacken\path(2280,3210)(1920,3300)(2280,3390)(2280,3210)
\blacken\path(2280,3810)(1920,3900)(2280,3990)(2280,3810)
%
%
\blacken\path(2820,2010)(3180,2100)(2820,2190)(2820,2010)
\blacken\path(1290,1860)(1200,1500)(1110,1860)(1290,1860)
\blacken\path(1890,1740)(1800,2100)(1710,1740)(1890,1740)
\blacken\path(2490,1740)(2400,2100)(2310,1740)(2490,1740)
\blacken\path(3690,1740)(3600,2100)(3510,1740)(3690,1740)
\blacken\path(4290,1740)(4200,2100)(4110,1740)(4290,1740)
\blacken\path(4890,1740)(4800,2100)(4710,1740)(4890,1740)
\blacken\path(5490,1740)(5400,2100)(5310,1740)(5490,1740)
\blacken\path(6090,1740)(6000,2100)(5910,1740)(6090,1740)
\blacken\path(6690,1740)(6600,2100)(6510,1740)(6690,1740)
\blacken\path(7290,1740)(7200,2100)(7110,1740)(7290,1740)
\blacken\path(7890,1740)(7800,2100)(7710,1740)(7890,1740)
\blacken\path(8490,1740)(8400,2100)(8310,1740)(8490,1740)
\blacken\path(1290,1860)(1200,1500)(1110,1860)(1290,1860)
\blacken\path(1290,2460)(1200,2100)(1110,2460)(1290,2460)
\blacken\path(1290,3060)(1200,2700)(1110,3060)(1290,3060)
\blacken\path(1290,3660)(1200,3300)(1110,3660)(1290,3660)
\blacken\path(1290,4560)(1200,4200)(1110,4560)(1290,4560)
%
%
\blacken\path(1890,2460)(1800,2100)(1710,2460)(1890,2460)
\blacken\path(1890,3060)(1800,2700)(1710,3060)(1890,3060)
\blacken\path(1890,3660)(1800,3300)(1710,3660)(1890,3660)
\blacken\path(1890,4560)(1800,4200)(1710,4560)(1890,4560)
%
%
\blacken\path(2490,2340)(2400,2700)(2310,2340)(2490,2340)
%
%
\blacken\path(3700,2340)(3612,2700)(3520,2340)(3700,2340)
\blacken\path(4300,2340)(4212,2700)(4120,2340)(4300,2340)
\blacken\path(4900,2340)(4812,2700)(4720,2340)(4900,2340)
\blacken\path(5500,2340)(5412,2700)(5320,2340)(5500,2340)
\blacken\path(6100,2340)(6012,2700)(5920,2340)(6100,2340)
\blacken\path(6700,2340)(6612,2700)(6520,2340)(6700,2340)
\blacken\path(7300,2340)(7212,2700)(7120,2340)(7300,2340)
\blacken\path(7900,2340)(7812,2700)(7720,2340)(7900,2340)
\blacken\path(8500,2340)(8412,2700)(8320,2340)(8500,2340)
%
\blacken\path(3720,2010)(4080,2100)(3720,2190)(3720,2010)
\blacken\path(4320,2010)(4680,2100)(4320,2190)(4320,2010)
\blacken\path(4920,2010)(5280,2100)(4920,2190)(4920,2010)
\blacken\path(5520,2010)(5880,2100)(5520,2190)(5520,2010)
\blacken\path(6120,2010)(6480,2100)(6120,2190)(6120,2010)
\blacken\path(6720,2010)(7080,2100)(6720,2190)(6720,2010)
\blacken\path(7320,2010)(7680,2100)(7320,2190)(7320,2010)
\blacken\path(7920,2010)(8280,2100)(7920,2190)(7920,2010)
%
\blacken\path(2820,2610)(3180,2700)(2820,2790)(2820,2610)
\blacken\path(3720,2610)(4080,2700)(3720,2790)(3720,2610)
\blacken\path(4320,2610)(4680,2700)(4320,2790)(4320,2610)
\blacken\path(4920,2610)(5280,2700)(4920,2790)(4920,2610)
\blacken\path(5520,2610)(5880,2700)(5520,2790)(5520,2610)
\blacken\path(6120,2610)(6480,2700)(6120,2790)(6120,2610)
\blacken\path(6720,2610)(7080,2700)(6720,2790)(6720,2610)
\blacken\path(7320,2610)(7680,2700)(7320,2790)(7320,2610)
\blacken\path(7920,2610)(8280,2700)(7920,2790)(7920,2610)
%
%
\blacken\path(2500,3060)(2410,2700)(2320,3060)(2500,3060)
\blacken\path(2500,3660)(2410,3300)(2320,3660)(2500,3660)
\blacken\path(2500,4560)(2410,4200)(2320,4560)(2500,4560)
%
%
\blacken\path(3180,3210)(2820,3300)(3180,3390)(3180,3210)
\blacken\path(3180,3810)(2820,3900)(3180,3990)(3180,3810)
\blacken\path(3700,2940)(3610,3300)(3520,2940)(3700,2940)
\blacken\path(4300,2940)(4210,3300)(4120,2940)(4300,2940)
\blacken\path(4900,2940)(4810,3300)(4720,2940)(4900,2940)
\blacken\path(5500,2940)(5410,3300)(5320,2940)(5500,2940)
\blacken\path(6100,2940)(6010,3300)(5920,2940)(6100,2940)
\blacken\path(6700,2940)(6610,3300)(6520,2940)(6700,2940)
\blacken\path(7300,2940)(7210,3300)(7120,2940)(7300,2940)
\blacken\path(7900,2940)(7810,3300)(7720,2940)(7900,2940)
\blacken\path(8500,2940)(8410,3300)(8320,2940)(8500,2940)
\end{picture}
%
\caption{The effect of forcing the three vertices at the intersection 
of the $\{v_1, z_1\}$, $\{v_2, z_2\}$ and $\{v_3, z_3\}$
rapidity lines to be a $c$-vertices. We used the 
notation $N_3 = N_1 - N_2$.}
%
\label{bc-3}
\end{figure}
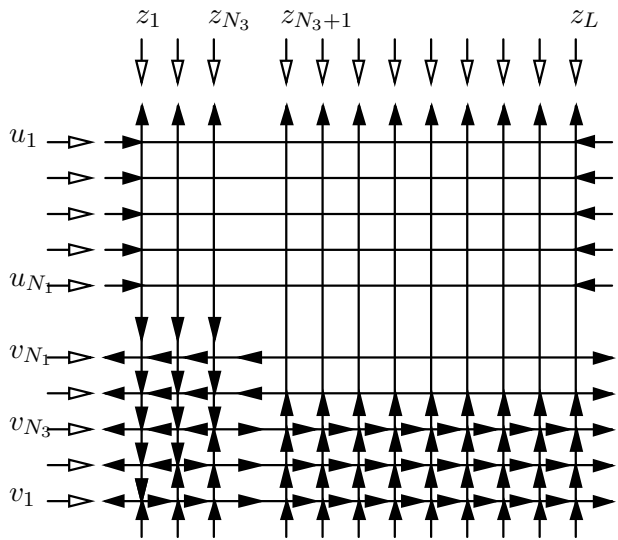

The above exercise in {\it `freezing'} vertices and arrows 
can be repeated and to produce a non-trivial example, we do 
it two more times. 
Setting $v_2 = z_2$ forces the vertex at the intersection 
of the lines carrying the rapidities $v_2$ and $z_2$ to be 
a $c$-vertex and freezes all arrows to the right as well as 
all arrows above that vertex and along $C$-lines. Setting 
$v_3 = z_3$, we end up with the lattice configuration 
in Figure {\bf \ref{bc-3}}.

%
\bigskip

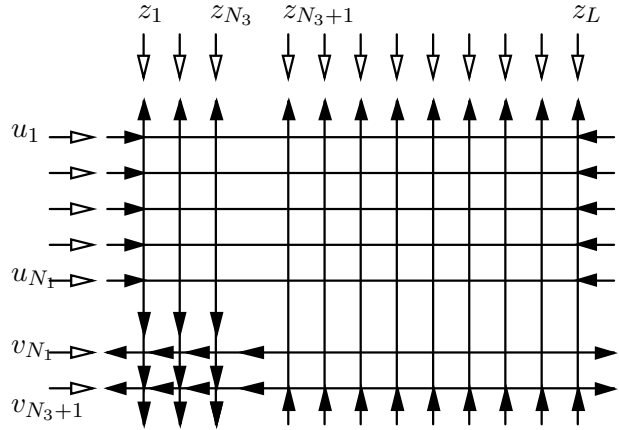
\begin{figure}
\setlength{\unitlength}{0.0008cm}
\begin{picture}(10000,07500)(0000,2500)
\thicklines
%
%
\path(0600,3300)(9000,3300)
\path(0600,3900)(9000,3900)
\path(0600,5100)(9000,5100)
\path(0600,5700)(9000,5700)
\path(0600,6300)(9000,6300)
\path(0600,6900)(9000,6900)
\path(0600,7500)(9000,7500)
%
\path(1200,8100)(1200,2700)
\path(1800,8100)(1800,2700)
\path(2400,8100)(2400,2700)
\path(3600,8100)(3600,2700)
\path(4200,8100)(4200,2700)
\path(4800,8100)(4800,2700)
\path(5400,8100)(5400,2700)
\path(6000,8100)(6000,2700)
\path(6600,8100)(6600,2700)
\path(7200,8100)(7200,2700)
\path(7800,8100)(7800,2700)
\path(8400,8100)(8400,2700)
%
%
\blacken\path(0900,3390)(0600,3300)(0900,3210)(0900,3390)
\blacken\path(0900,3990)(0600,3900)(0900,3810)(0900,3990)
\blacken\path(0900,5010)(1200,5110)(0900,5190)(0900,5010)
\blacken\path(0900,5610)(1200,5700)(0900,5790)(0900,5610)
\blacken\path(0900,6210)(1210,6300)(0900,6390)(0900,6210)
\blacken\path(0900,6810)(1200,6900)(0900,6990)(0900,6810)
\blacken\path(0900,7410)(1200,7500)(0900,7590)(0900,7410)
\blacken\path(1300,7800)(1210,8100)(1120,7800)(1300,7800)
\blacken\path(1900,7800)(1810,8100)(1720,7800)(1900,7800)
\blacken\path(2500,7800)(2410,8100)(2320,7800)(2500,7800)
\blacken\path(3700,7800)(3610,8100)(3520,7800)(3700,7800)
\blacken\path(4300,7800)(4210,8100)(4120,7800)(4300,7800)
\blacken\path(4900,7800)(4810,8100)(4720,7800)(4900,7800)
\blacken\path(5500,7800)(5410,8100)(5320,7800)(5500,7800)
\blacken\path(6100,7800)(6010,8100)(5920,7800)(6100,7800)
\blacken\path(6700,7800)(6610,8100)(6520,7800)(6700,7800)
\blacken\path(7300,7800)(7210,8100)(7120,7800)(7300,7800)
\blacken\path(7900,7800)(7810,8100)(7720,7800)(7900,7800)
\blacken\path(8500,7800)(8410,8100)(8320,7800)(8500,7800)
\blacken\path(8700,3200)(9000,3290)(8700,3380)(8700,3200)
\blacken\path(8700,3800)(9000,3890)(8700,3980)(8700,3800)
\blacken\path(8700,5200)(8400,5110)(8700,5020)(8700,5200)
\blacken\path(8700,5800)(8400,5710)(8700,5620)(8700,5800)
\blacken\path(8700,6400)(8400,6310)(8700,6220)(8700,6400)
\blacken\path(8700,7000)(8400,6910)(8700,6820)(8700,7000)
\blacken\path(8700,7600)(8400,7510)(8700,7420)(8700,7600)
%
%
\path(-0360,3300)(0000,3300)
\path(-0360,3900)(0000,3900)
\put(-1000,3900){$v_{ N_1          }$}
\put(-1000,2900){$v_{N_3 + 1}$}
\whiten\path(0000,3210)(0360,3300)(0000,3390)(0000,3210)
\whiten\path(0000,3810)(0360,3900)(0000,3990)(0000,3810)
\path(-0360,5100)(0000,5100)
\path(-0360,5700)(0000,5700)
\path(-0360,6300)(0000,6300)
\path(-0360,6900)(0000,6900)
\path(-0360,7500)(0000,7500)
\put(-1000,5100){$u_{N_1}$}
\put(-1000,7500){$u_1$}
\whiten\path(0000,5010)(0360,5100)(0000,5190)(0000,5010)
\whiten\path(0000,5610)(0360,5700)(0000,5790)(0000,5610)
\whiten\path(0000,6210)(0360,6300)(0000,6390)(0000,6210)
\whiten\path(0000,6810)(0360,6900)(0000,6990)(0000,6810)
\whiten\path(0000,7410)(0360,7500)(0000,7590)(0000,7410)
%
%
\path(1200,9220)(1200,8500)
\path(1800,9220)(1800,8500)
\path(2400,9220)(2400,8500)
\path(3600,9220)(3600,8500)
\path(4200,9220)(4200,8500)
\path(4800,9220)(4800,8500)
\path(5400,9220)(5400,8500)
\path(6000,9220)(6000,8500)
\path(6600,9220)(6600,8500)
\path(7200,9220)(7200,8500)
\path(7800,9220)(7800,8500)
\path(8400,9220)(8400,8500)
\put(1100,9500){$z_{      1}$}
\put(2300,9500){$z_{N_3    }$}
\put(3500,9500){$z_{N_3 + 1}$}
\put(8300,9500){$z_{L      }$}
\whiten\path(1290,8860)(1200,8500)(1110,8860)(1290,8860)
\whiten\path(1890,8860)(1800,8500)(1710,8860)(1890,8860)
\whiten\path(2490,8860)(2400,8500)(2310,8860)(2490,8860)
\whiten\path(3690,8860)(3600,8500)(3510,8860)(3690,8860)
\whiten\path(4290,8860)(4200,8500)(4110,8860)(4290,8860)
\whiten\path(4890,8860)(4800,8500)(4710,8860)(4890,8860)
\whiten\path(5490,8860)(5400,8500)(5310,8860)(5490,8860)
\whiten\path(6090,8860)(6000,8500)(5910,8860)(6090,8860)
\whiten\path(6690,8860)(6600,8500)(6510,8860)(6690,8860)
\whiten\path(7290,8860)(7200,8500)(7110,8860)(7290,8860)
\whiten\path(7890,8860)(7800,8500)(7710,8860)(7890,8860)
\whiten\path(8490,8860)(8400,8500)(8310,8860)(8490,8860)
%
%
%
\blacken\path(1680,3210)(1320,3300)(1680,3390)(1680,3210)
\blacken\path(1680,3810)(1320,3900)(1680,3990)(1680,3810)
%
%
\blacken\path(2280,3210)(1920,3300)(2280,3390)(2280,3210)
\blacken\path(2280,3810)(1920,3900)(2280,3990)(2280,3810)
%
%
%
%
\blacken\path(1890,3660)(1800,3300)(1710,3660)(1890,3660)
\blacken\path(1890,4560)(1800,4200)(1710,4560)(1890,4560)
%
%
\blacken\path(2500,3060)(2410,2700)(2320,3060)(2500,3060)
%
%
\blacken\path(2500,3660)(2410,3300)(2320,3660)(2500,3660)
\blacken\path(2500,4560)(2410,4200)(2320,4560)(2500,4560)
%
%
\blacken\path(3180,3210)(2820,3300)(3180,3390)(3180,3210)
\blacken\path(3180,3810)(2820,3900)(3180,3990)(3180,3810)
%
%
\blacken\path(1300,3060)(1210,2700)(1120,3060)(1300,3060)
\blacken\path(1900,3060)(1810,2700)(1720,3060)(1900,3060)
\blacken\path(2500,3060)(2410,2700)(2320,3060)(2500,3060)
%
\blacken\path(1300,3660)(1210,3300)(1120,3660)(1300,3660)
\blacken\path(1300,4560)(1210,4200)(1120,4560)(1300,4560)
%
\blacken\path(3700,2940)(3610,3300)(3520,2940)(3700,2940)
\blacken\path(4300,2940)(4210,3300)(4120,2940)(4300,2940)
\blacken\path(4900,2940)(4810,3300)(4720,2940)(4900,2940)
\blacken\path(5500,2940)(5410,3300)(5320,2940)(5500,2940)
\blacken\path(6100,2940)(6010,3300)(5920,2940)(6100,2940)
\blacken\path(6700,2940)(6610,3300)(6520,2940)(6700,2940)
\blacken\path(7300,2940)(7210,3300)(7120,2940)(7300,2940)
\blacken\path(7900,2940)(7810,3300)(7720,2940)(7900,2940)
\blacken\path(8500,2940)(8410,3300)(8320,2940)(8500,2940)
\end{picture}
%
\caption{A restricted $[L, N_1, N_2]$-configuration. 
In this example, 
$N_1=5$, $N_2 = 2$, and as always $N_3 = N_1 - N_2$.}
%
\label{restricted-bc}
\end{figure}

From Figure {\bf {\ref{bc-3}}}, one can see that 
\1 All arrows on the lower $N_3$ horizontal lines, 
where $N_3=3$ in the specific example shown, are 
frozen, and 
\2 All lines on the $N_3$ left most vertical lines 
in the lower half of the diagram, where they 
intersect with $C$-lines. Removing the lower 
$N_3$ $C$-lines we obtain the 
configuration in Figure {\bf \ref{restricted-bc}}.
This configuration has a subset (rectangular shape 
on lower left corner) that is also completely frozen. 
All vertices in this part are $a$-vertices, hence 
from Equation {\bf \ref{weights}}, their contribution 
to the partition function of this configuration is 
trivial.

An $[L, N_1, N_2]$-configuration, as in 
Figure {\bf \ref{restricted-bc}}, interpolates between 
an initial reference state 
$| [L^{\wedge}] \>$
and a final 
$ \< [{N_3}^{\vee}, (L-N_3)^{\wedge}] |$ 
state, using $N_1$ $B$-lines followed by $N_2$ $C$-lines.

%
\bigskip

%
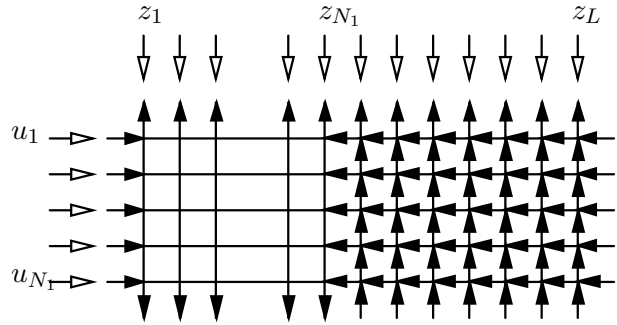
\begin{figure}
\setlength{\unitlength}{0.0008cm}
\begin{picture}(10000,06000)(0000,4000)
\thicklines
%
%
\path(0600,5100)(9000,5100)
\path(0600,5700)(9000,5700)
\path(0600,6300)(9000,6300)
\path(0600,6900)(9000,6900)
\path(0600,7500)(9000,7500)
%
\path(1200,8100)(1200,4500)
\path(1800,8100)(1800,4500)
\path(2400,8100)(2400,4500)
\path(3600,8100)(3600,4500)
\path(4200,8100)(4200,4500)
\path(4800,8100)(4800,4500)
\path(5400,8100)(5400,4500)
\path(6000,8100)(6000,4500)
\path(6600,8100)(6600,4500)
\path(7200,8100)(7200,4500)
\path(7800,8100)(7800,4500)
\path(8400,8100)(8400,4500)
%
\blacken\path(0900,5010)(1200,5100)(0900,5190)(0900,5010)
\blacken\path(0900,5610)(1200,5700)(0900,5790)(0900,5610)
\blacken\path(0900,6210)(1210,6300)(0900,6390)(0900,6210)
\blacken\path(0900,6810)(1200,6900)(0900,6990)(0900,6810)
\blacken\path(0900,7410)(1200,7500)(0900,7590)(0900,7410)
\blacken\path(1300,7800)(1210,8100)(1120,7800)(1300,7800)
\blacken\path(1900,7800)(1810,8100)(1720,7800)(1900,7800)
\blacken\path(2500,7800)(2410,8100)(2320,7800)(2500,7800)
\blacken\path(3700,7800)(3610,8100)(3520,7800)(3700,7800)
\blacken\path(4300,7800)(4210,8100)(4120,7800)(4300,7800)
\blacken\path(4900,7800)(4810,8100)(4720,7800)(4900,7800)
\blacken\path(5500,7800)(5410,8100)(5320,7800)(5500,7800)
\blacken\path(6100,7800)(6010,8100)(5920,7800)(6100,7800)
\blacken\path(6700,7800)(6610,8100)(6520,7800)(6700,7800)
\blacken\path(7300,7800)(7210,8100)(7120,7800)(7300,7800)
\blacken\path(7900,7800)(7810,8100)(7720,7800)(7900,7800)
\blacken\path(8500,7800)(8410,8100)(8320,7800)(8500,7800)
%
%
\blacken\path(8760,5200)(8400,5110)(8760,5020)(8760,5200)
\blacken\path(8160,5200)(7800,5110)(8160,5020)(8160,5200)
\blacken\path(7560,5200)(7200,5110)(7560,5020)(7560,5200)
\blacken\path(6960,5200)(6600,5110)(6960,5020)(6960,5200)
\blacken\path(6360,5200)(6000,5110)(6360,5020)(6360,5200)
\blacken\path(5760,5200)(5400,5110)(5760,5020)(5760,5200)
\blacken\path(5160,5200)(4800,5110)(5160,5020)(5160,5200)
\blacken\path(4560,5200)(4200,5110)(4560,5020)(4560,5200)
%
%
%
%
\blacken\path(4560,5800)(4200,5710)(4560,5620)(4560,5800)
\blacken\path(5160,5800)(4800,5710)(5160,5620)(5160,5800)
\blacken\path(5760,5800)(5400,5710)(5760,5620)(5760,5800)
\blacken\path(6360,5800)(6000,5710)(6360,5620)(6360,5800)
\blacken\path(6960,5800)(6600,5710)(6960,5620)(6960,5800)
\blacken\path(7560,5800)(7200,5710)(7560,5620)(7560,5800)
\blacken\path(8160,5800)(7800,5710)(8160,5620)(8160,5800)
%
%
\blacken\path(4560,6400)(4200,6310)(4560,6220)(4560,6400)
\blacken\path(5160,6400)(4800,6310)(5160,6220)(5160,6400)
\blacken\path(5760,6400)(5400,6310)(5760,6220)(5760,6400)
\blacken\path(6360,6400)(6000,6310)(6360,6220)(6360,6400)
\blacken\path(6960,6400)(6600,6310)(6960,6220)(6960,6400)
\blacken\path(7560,6400)(7200,6310)(7560,6220)(7560,6400)
\blacken\path(8160,6400)(7800,6310)(8160,6220)(8160,6400)
%
%
\blacken\path(4560,7000)(4200,6910)(4560,6820)(4560,7000)
\blacken\path(5160,7000)(4800,6910)(5160,6820)(5160,7000)
\blacken\path(5760,7000)(5400,6910)(5760,6820)(5760,7000)
\blacken\path(6360,7000)(6000,6910)(6360,6820)(6360,7000)
\blacken\path(6960,7000)(6600,6910)(6960,6820)(6960,7000)
\blacken\path(7560,7000)(7200,6910)(7560,6820)(7560,7000)
\blacken\path(8160,7000)(7800,6910)(8160,6820)(8160,7000)
%
%
\blacken\path(4560,7600)(4200,7510)(4560,7420)(4560,7600)
\blacken\path(5160,7600)(4800,7510)(5160,7420)(5160,7600)
\blacken\path(5760,7600)(5400,7510)(5760,7420)(5760,7600)
\blacken\path(6360,7600)(6000,7510)(6360,7420)(6360,7600)
\blacken\path(6960,7600)(6600,7510)(6960,7420)(6960,7600)
\blacken\path(7560,7600)(7200,7510)(7560,7420)(7560,7600)
\blacken\path(8160,7600)(7800,7510)(8160,7420)(8160,7600)
\blacken\path(8700,5800)(8400,5710)(8700,5620)(8700,5800)
\blacken\path(8700,6400)(8400,6310)(8700,6220)(8700,6400)
\blacken\path(8700,7000)(8400,6910)(8700,6820)(8700,7000)
\blacken\path(8700,7600)(8400,7510)(8700,7420)(8700,7600)
%
%
\path(-0360,5100)(0000,5100)
\path(-0360,5700)(0000,5700)
\path(-0360,6300)(0000,6300)
\path(-0360,6900)(0000,6900)
\path(-0360,7500)(0000,7500)
\put(-1000,5100){$u_{N_1}$}
\put(-1000,7500){$u_{  1}$}
\whiten\path(0000,5010)(0360,5100)(0000,5190)(0000,5010)
\whiten\path(0000,5610)(0360,5700)(0000,5790)(0000,5610)
\whiten\path(0000,6210)(0360,6300)(0000,6390)(0000,6210)
\whiten\path(0000,6810)(0360,6900)(0000,6990)(0000,6810)
\whiten\path(0000,7410)(0360,7500)(0000,7590)(0000,7410)
%
%
\path(1200,9220)(1200,8500)
\path(1800,9220)(1800,8500)
\path(2400,9220)(2400,8500)
\path(3600,9220)(3600,8500)
\path(4200,9220)(4200,8500)
\path(4800,9220)(4800,8500)
\path(5400,9220)(5400,8500)
\path(6000,9220)(6000,8500)
\path(6600,9220)(6600,8500)
\path(7200,9220)(7200,8500)
\path(7800,9220)(7800,8500)
\path(8400,9220)(8400,8500)
\put(1100,9500){$z_{     1}$}
\put(4100,9500){$z_{N_1   }$}
\put(8300,9500){$z_{L}$}
\whiten\path(1290,8860)(1200,8500)(1110,8860)(1290,8860)
\whiten\path(1890,8860)(1800,8500)(1710,8860)(1890,8860)
\whiten\path(2490,8860)(2400,8500)(2310,8860)(2490,8860)
\whiten\path(3690,8860)(3600,8500)(3510,8860)(3690,8860)
\whiten\path(4290,8860)(4200,8500)(4110,8860)(4290,8860)
\whiten\path(4890,8860)(4800,8500)(4710,8860)(4890,8860)
\whiten\path(5490,8860)(5400,8500)(5310,8860)(5490,8860)
\whiten\path(6090,8860)(6000,8500)(5910,8860)(6090,8860)
\whiten\path(6690,8860)(6600,8500)(6510,8860)(6690,8860)
\whiten\path(7290,8860)(7200,8500)(7110,8860)(7290,8860)
\whiten\path(7890,8860)(7800,8500)(7710,8860)(7890,8860)
\whiten\path(8490,8860)(8400,8500)(8310,8860)(8490,8860)
%
%
\blacken\path(1300,4860)(1210,4500)(1120,4860)(1300,4860)
\blacken\path(1900,4860)(1810,4500)(1720,4860)(1900,4860)
\blacken\path(2500,4860)(2410,4500)(2320,4860)(2500,4860)
\blacken\path(3700,4860)(3610,4500)(3520,4860)(3700,4860)
\blacken\path(4300,4860)(4210,4500)(4120,4860)(4300,4860)
%
%
%
\blacken\path(4900,4740)(4810,5100)(4720,4740)(4900,4740)
\blacken\path(4900,5340)(4810,5700)(4720,5340)(4900,5340)
\blacken\path(4900,5940)(4810,6300)(4720,5940)(4900,5940)
\blacken\path(4900,6540)(4810,6900)(4720,6540)(4900,6540)
\blacken\path(4900,7140)(4810,7500)(4720,7140)(4900,7140)
%
%
\blacken\path(5500,4740)(5410,5100)(5320,4740)(5500,4740)
\blacken\path(5500,5340)(5410,5700)(5320,5340)(5500,5340)
\blacken\path(5500,5940)(5410,6300)(5320,5940)(5500,5940)
\blacken\path(5500,6540)(5410,6900)(5320,6540)(5500,6540)
\blacken\path(5500,7140)(5410,7500)(5320,7140)(5500,7140)
%
%
\blacken\path(6100,4740)(6010,5100)(5920,4740)(6100,4740)
\blacken\path(6100,5340)(6010,5700)(5920,5340)(6100,5340)
\blacken\path(6100,5940)(6010,6300)(5920,5940)(6100,5940)
\blacken\path(6100,6540)(6010,6900)(5920,6540)(6100,6540)
\blacken\path(6100,7140)(6010,7500)(5920,7140)(6100,7140)
%
%
\blacken\path(6700,4740)(6610,5100)(6520,4740)(6700,4740)
\blacken\path(6700,5340)(6610,5700)(6520,5340)(6700,5340)
\blacken\path(6700,5940)(6610,6300)(6520,5940)(6700,5940)
\blacken\path(6700,6540)(6610,6900)(6520,6540)(6700,6540)
\blacken\path(6700,7140)(6610,7500)(6520,7140)(6700,7140)
%
%
\blacken\path(7300,4740)(7210,5100)(7120,4740)(7300,4740)
\blacken\path(7300,5340)(7210,5700)(7120,5340)(7300,5340)
\blacken\path(7300,5940)(7210,6300)(7120,5940)(7300,5940)
\blacken\path(7300,6540)(7210,6900)(7120,6540)(7300,6540)
\blacken\path(7300,7140)(7210,7500)(7120,7140)(7300,7140)
%
%
\blacken\path(7900,4740)(7810,5100)(7720,4740)(7900,4740)
\blacken\path(7900,5340)(7810,5700)(7720,5340)(7900,5340)
\blacken\path(7900,5940)(7810,6300)(7720,5940)(7900,5940)
\blacken\path(7900,6540)(7810,6900)(7720,6540)(7900,6540)
\blacken\path(7900,7140)(7810,7500)(7720,7140)(7900,7140)
%
%
\blacken\path(8500,4740)(8410,5100)(8320,4740)(8500,4740)
\blacken\path(8500,5340)(8410,5700)(8320,5340)(8500,5340)
\blacken\path(8500,5940)(8410,6300)(8320,5940)(8500,5940)
\blacken\path(8500,6540)(8410,6900)(8320,6540)(8500,6540)
\blacken\path(8500,7140)(8410,7500)(8320,7140)(8500,7140)
\end{picture}
%
\caption{A restricted $[L, N_1, N_2]$-configuration. In 
this example, $N_1=5$ and $N_2=5$. Equivalently, the left 
half is an $(N_1 \! \times \! N_1)$ 
domain wall configuration, where $N_1=5$, with an additional 
totally frozen lattice configuration to its right.}
%
%
\label{dw-extended}
\end{figure}
\bigskip

Setting $v_i = z_i$ for $i = 1, \cdots, N_1$, we freeze all arrows 
that are on $C$-lines or on segments that end on $C$-lines. 
Discarding these we obtain the lattice configuration in Figure 
{\bf \ref{dw-extended}}.

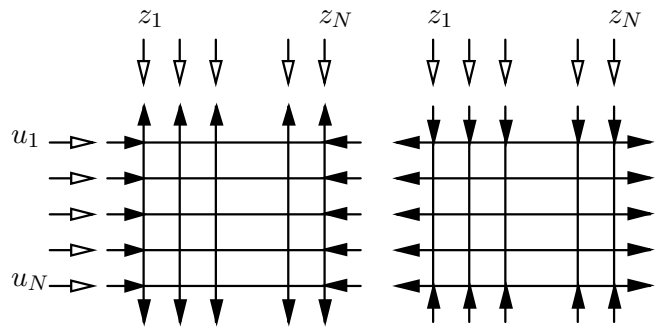
\begin{figure}
\setlength{\unitlength}{0.0008cm}
\begin{picture}(10000,06000)(0000,4000)
\thicklines
%
%
\path(0600,5100)(4800,5100)
\path(0600,5700)(4800,5700)
\path(0600,6300)(4800,6300)
\path(0600,6900)(4800,6900)
\path(0600,7500)(4800,7500)
%
\path(1200,8100)(1200,4500)
\path(1800,8100)(1800,4500)
\path(2400,8100)(2400,4500)
\path(3600,8100)(3600,4500)
\path(4200,8100)(4200,4500)
%
\blacken\path(0900,5010)(1200,5100)(0900,5190)(0900,5010)
\blacken\path(0900,5610)(1200,5700)(0900,5790)(0900,5610)
\blacken\path(0900,6210)(1210,6300)(0900,6390)(0900,6210)
\blacken\path(0900,6810)(1200,6900)(0900,6990)(0900,6810)
\blacken\path(0900,7410)(1200,7500)(0900,7590)(0900,7410)
\blacken\path(1300,7800)(1210,8100)(1120,7800)(1300,7800)
\blacken\path(1900,7800)(1810,8100)(1720,7800)(1900,7800)
\blacken\path(2500,7800)(2410,8100)(2320,7800)(2500,7800)
\blacken\path(3700,7800)(3610,8100)(3520,7800)(3700,7800)
\blacken\path(4300,7800)(4210,8100)(4120,7800)(4300,7800)
\blacken\path(4560,5200)(4200,5110)(4560,5020)(4560,5200)
%
%
\blacken\path(4560,5800)(4200,5710)(4560,5620)(4560,5800)
\blacken\path(4560,6400)(4200,6310)(4560,6220)(4560,6400)
\blacken\path(4560,7000)(4200,6910)(4560,6820)(4560,7000)
\blacken\path(4560,7600)(4200,7510)(4560,7420)(4560,7600)
%
%
\path(-0360,5100)(0000,5100)
\path(-0360,5700)(0000,5700)
\path(-0360,6300)(0000,6300)
\path(-0360,6900)(0000,6900)
\path(-0360,7500)(0000,7500)
\put(-1000,5100){$u_N$}
\put(-1000,7500){$u_1$}
\whiten\path(0000,5010)(0360,5100)(0000,5190)(0000,5010)
\whiten\path(0000,5610)(0360,5700)(0000,5790)(0000,5610)
\whiten\path(0000,6210)(0360,6300)(0000,6390)(0000,6210)
\whiten\path(0000,6810)(0360,6900)(0000,6990)(0000,6810)
\whiten\path(0000,7410)(0360,7500)(0000,7590)(0000,7410)
%
%
\path(1200,9220)(1200,8500)
\path(1800,9220)(1800,8500)
\path(2400,9220)(2400,8500)
\path(3600,9220)(3600,8500)
\path(4200,9220)(4200,8500)
\put(1100,9500){$z_1   $}
\put(4100,9500){$z_N   $}
\whiten\path(1290,8860)(1200,8500)(1110,8860)(1290,8860)
\whiten\path(1890,8860)(1800,8500)(1710,8860)(1890,8860)
\whiten\path(2490,8860)(2400,8500)(2310,8860)(2490,8860)
\whiten\path(3690,8860)(3600,8500)(3510,8860)(3690,8860)
\whiten\path(4290,8860)(4200,8500)(4110,8860)(4290,8860)
%
%
\blacken\path(1300,4860)(1210,4500)(1120,4860)(1300,4860)
\blacken\path(1900,4860)(1810,4500)(1720,4860)(1900,4860)
\blacken\path(2500,4860)(2410,4500)(2320,4860)(2500,4860)
\blacken\path(3700,4860)(3610,4500)(3520,4860)(3700,4860)
\blacken\path(4300,4860)(4210,4500)(4120,4860)(4300,4860)
%
%
%
%
%
\path(5400,7500)(9600,7500)
\path(5400,6900)(9600,6900)
\path(5400,6300)(9600,6300)
\path(5400,5700)(9600,5700)
\path(5400,5100)(9600,5100)
%
%
\path(6000,8100)(6000,4500)
\path(6600,8100)(6600,4500)
\path(7200,8100)(7200,4500)
\path(8400,8100)(8400,4500)
\path(9000,8100)(9000,4500)
%
%
\blacken\path(5760,5010)(5400,5100)(5760,5190)(5760,5010)
\blacken\path(5760,5610)(5400,5700)(5760,5790)(5760,5610)
\blacken\path(5760,6210)(5400,6300)(5760,6390)(5760,6210)
\blacken\path(5760,6810)(5400,6900)(5760,6990)(5760,6810)
\blacken\path(5760,7410)(5400,7500)(5760,7590)(5760,7410)
%
%
\blacken\path(6100,7860)(6010,7500)(5920,7860)(6100,7860)
\blacken\path(6700,7860)(6610,7500)(6520,7860)(6700,7860)
\blacken\path(7300,7860)(7210,7500)(7120,7860)(7300,7860)
\blacken\path(8500,7860)(8410,7500)(8320,7860)(8500,7860)
\blacken\path(9100,7860)(9010,7500)(8920,7860)(9100,7860)
%
%
\blacken\path(9240,7600)(9600,7510)(9240,7420)(9240,7600)
\blacken\path(9240,7000)(9600,6910)(9240,6820)(9240,7000)
\blacken\path(9240,6400)(9600,6310)(9240,6220)(9240,6400)
\blacken\path(9240,5800)(9600,5710)(9240,5620)(9240,5800)
\blacken\path(9240,5200)(9600,5110)(9240,5020)(9240,5200)
%
%
\path(6000,9220)(6000,8500)
\path(6600,9220)(6600,8500)
\path(7200,9220)(7200,8500)
\path(8400,9220)(8400,8500)
\path(9000,9220)(9000,8500)
\put(5900,9500){$z_1   $}
\put(8900,9500){$z_N   $}
\whiten\path(6090,8860)(6000,8500)(5910,8860)(6090,8860)
\whiten\path(6690,8860)(6600,8500)(6510,8860)(6690,8860)
\whiten\path(7290,8860)(7200,8500)(7110,8860)(7290,8860)
\whiten\path(8490,8860)(8400,8500)(8310,8860)(8490,8860)
\whiten\path(9090,8860)(9000,8500)(8910,8860)(9090,8860)
%
%
\blacken\path(6100,4740)(6010,5100)(5920,4740)(6100,4740)
\blacken\path(6700,4740)(6610,5100)(6520,4740)(6700,4740)
\blacken\path(7300,4740)(7210,5100)(7120,4740)(7300,4740)
\blacken\path(8500,4740)(8410,5100)(8320,4740)(8500,4740)
\blacken\path(9100,4740)(9010,5100)(8920,4740)(9100,4740)
\end{picture}
%
\caption{The left hand side is an $(N \! \times  \! N)$ domain wall 
configuration, where $N=5$. The right hand side is the 
corresponding dual configuration.}
%
\label{dw}
\end{figure}

Removing all frozen vertices (as well as the extra space 
between two sets of vertical lines, that is no longer
necessary), one obtains the {\it domain wall configuration} 
in Figure {\bf \ref{dw}}, which is characterized as 
follows. All arrows on the left and right boundaries point 
inwards, and all arrows on the upper and lower boundaries 
point outwards. The internal arrows remain free, and the
configurations that are consistent with the boundary conditions
are summed over. Reversing the orientation of all arrows on all 
boundaries is a dual a domain wall configuration. 

%
\bigskip

\subsection{Remarks on domain wall configurations}
\1 One can generate a domain wall configuration directly 
starting from a length-$N$ initial reference state followed 
by $N$ $B$-lines. 
\2 One can generate a dual domain wall configuration directly 
starting from a length-$N$ dual initial reference state 
followed by $N$ $C$-lines.
\3 A $BC$-configuration with length-$L$ initial and final 
reference states, 
$L$ $B$-lines and $L$ $C$-lines, factorizes into a product 
of 
a domain wall configuration and 
a dual domain wall configuration.
\4 The restriction of $BC$-configurations to 
$[L, N_1, N_2]$-configurations, where $N_2 < N_1$, 
produces 
a recursion relation that was used by Wheeler in 
\cite{MW} to provide a recursive proof of Slavnov's determinant 
expression for the scalar product of a Bethe eigenstate and 
a generic state in the corresponding spin chain. 
\5 The partition function of a domain wall configuration has 
a determinant expression found by Izergin, that can be derived 
in six-vertex model terms (without reference to spin chains or 
the BA) \cite{Izergin}. 

\subsection{Izergin's expression for the domain wall partition 
function}

Let $\{w\}_N$ $=$  $\{w_1, \cdots,$ $ w_N\}$ and $\{z\}_N$ $=$ 
$\{z_1, \cdots, z_N\}$ be two sets of variables\footnote{The 
following result does not require that any set 
of rapidities satisfy Bethe equations.}. Izergin's determinant 
expression for the domain wall partition function is 

\begin{multline}
Z_N \ll \{w\}_N,\{z\}_N \rr
=
\\
\frac{
\prod_{i, j=1}^{N} 
(w_i - z_j + \eta)
}
{
\prod_{1 \leq i < j \leq N} 
(w_i-w_j) (z_j-z_i)
}
\det
\ll
\frac{
\eta
}
{
(w_i - z_j + \eta) (w_i - z_j)
}
\rr_{1\leq i,j \leq N}
\label{dwpf}
\end{multline}

Dual domain wall configurations have the same partition 
functions due to invariance under reversing all arrows.
For the result of this note, we need 
the homogeneous limit of the above expression. Taking 
the limit $z_i \rightarrow z$, $\{i = 1, \cdots, L\}$, 
we obtain 

\begin{multline}
Z_{N}^{\textit{hom}} \ll \{w\}_N, z \rr
=
\frac{
\prod_{i}^{N}
(w_i - z + \eta)^{N}
}
{
\prod_{1 \leq i < j \leq N} 
(w_i - w_j) 
}
\det
\ll
\phi^{(j-1)}(w_i, z)
\rr_{1\leq i,j \leq N}
\\ 
\phi^{(j)}(w_i, z) = \frac{1}{j!}  
\  
\partial^{(j)}_z
\ll
\frac{
\eta
}{
(w_i - z + \eta) (w_i -  z)
}
\rr
\label{izergin-homogeneous}
\end{multline}

\section{The XXX spin-$\frac{1}{2}$ chain}
\label{section-spin-chain}

In this section, we recall the XXX spin-$\frac{1}{2}$ chain 
that we need to discuss the EGSV expression of the structure 
constants in \cite{E1}. Our aim is to motivate the connection 
with the rational six-vertex model discussed in Section 
{\bf \ref{section-six-vertex}}. 

\subsection{Closed spin chains, open lattice segments, 
and spin variables}
Consider a length-$L$ closed spin chain. Label the sites 
sequentially using $i \in \{1, 2, \cdots, $ $L\}$ and represent 
the closed spin chain as a length-$L$ segment of a 1-dimensional 
open lattice. Assign site $i$ a spin variable $\sigma_i$, 
$\sigma_1 \equiv \sigma_{L+1}$, that takes one of two possible 
values in a 2-dimensional space $h_i$ with a basis
$\ll \begin{array}{c} 1 \\ 0 \end{array} \rr_i$, 
$\ll \begin{array}{c} 0 \\ 1 \end{array} \rr_i$ 
which we refer to as `up' and `down'. 
The space of states ${\mathcal{H}}$ is the tensor product 
$\mathcal{H}  = h_1 \otimes \cdots $ $\otimes \ h_L$.
Every state in ${\mathcal{H}}$ is an assignment of $L$ 
definite-value (either up or down) spin variables to the 
sites of the spin chain. 
In computing scalar products, we wish to think of states 
in  ${\mathcal{H}}$ as initial states. 

\subsection{Initial reference and dual reference states}
${\mathcal{H}}$ contains two distinguished states, 
\begin{align}
|[ L^{\wedge} ] \> = \bigotimes_{i=1}^{L} 
\ll
\begin{array}{c}
1 \\ 0
\end{array}
\rr_i
,
\quad
|[L^{\vee} ]\> = \bigotimes_{i=1}^{L} 
\ll
\begin{array}{c}
0 \\ 1
\end{array}
\rr_i
\end{align}
\noindent where 
$[L^{\wedge}]$ indicates $L$ spin states all of which are up, and 
$[L^{\vee  }]$ indicates $L$ spin states all of which are down. 
These are the reference state and the dual reference state. 

\subsection{Final reference and dual reference states, 
and a variation}
Consider a length-$L$ spin chain, and assign each site $i$ 
the space $h_i^{*}$ with the basis 
$\ll 1 \ \ 0 \rr_i$, 
$\ll 0 \ \ 1 \rr_i$.
We construct a final space of states as the tensor product
${\mathcal{H}}^{*} = h_1^{*} \otimes \cdots $ $\otimes h_L^{*}$.
$\mathcal{H}^{*}$ contains two distinguished states
\begin{align}
\< [ L^{\wedge} ] | = \bigotimes_{i=1}^{L} 
\ll
\begin{array}{cc}
1 & 0
\end{array}
\rr_i
\quad
,
\quad 
\< [ L^{\vee}] | = \bigotimes_{i=1}^{L} 
\ll
\begin{array}{cc}
0 & 1
\end{array}
\rr_i
\end{align}
\noindent where all spins are up, and all spins are down. 
Finally, we consider the state
\begin{equation}
\< [{N_3}^{\vee}, (L-N_3)^{\wedge}] |
=
\bigotimes_{1 \leq i \leq N_3}
\ll
\begin{array}{cc}
0 & 1
\end{array}
\rr_i
\bigotimes_{(N_3 + 1) \leq i \leq L}
\ll
\begin{array}{cc}
1 & 0
\end{array}
\rr_i
\end{equation}

\noindent where first $N_3$ spins from the left are down, and all 
remaining spins up. 

\subsection{Remark} The connection to the six-vertex model is 
clear. Every state of the periodic spin chain is analogous to 
a spin set in the six-vertex model. Periodicity is not manifest 
in the latter representation for the same reason that it is not 
manifest once we choose a labeling system.  
The initial and final reference and dual reference states are 
the spin-chain analogues of those discussed in 
Section {\bf {\ref{section-six-vertex}}}.

\subsection{The $R$-matrix}

From an initial reference state, we can generate all other states 
in $\mathcal{H}$ using operators that flip the spin variables, one 
spin at a time. Defining these operators requires defining a sequence
of objects.
\1 The $R$-matrix,  
\2 The $L$-matrix, and finally,
\3 The monodromy or $M$-matrix.

The $R$-matrix assigns a weight to the transition from a pair 
of initial spin states (for example the definite spin states 
on the left and lower segments that meet at a certain vertex) 
to a pair of final spin states (the definite spin states on 
the right and upper segments that meet at the same vertex as 
the initial ones). In the case of the rational XXX 
spin-$\frac{1}{2}$, this a transition between four possible 
initial spin states and four final spin states and the 
$R$-matrix is the $(4 \! \times  \! 4)$-matrix
\begin{align}
R_{ab}(u_a, u_b)
=
\ll
\begin{array}{cccc}
a[u_a, u_b]     & 0              & 0            & 0    \\
0               & b[u_a, u_b]    & c[u_a, u_b]  & 0    \\
0               & c[u_a, u_b]    & b[u_a, u_b]  & 0    \\
0               & 0              & 0            & a[u_a, u_b]
\end{array}
\rr_{ab}
\label{R-matrix}
\end{align}

More formally, the $R$-matrix is an element of 
$\textit{End}$$(h_a\otimes h_b)$, where $h_a$ is an auxiliary 
space and $h_b$ is another auxiliary space of the spin chain. 
The variables $u_a, u_b$ are the corresponding rapidity variables. 
The $R$-matrix intertwines these spaces.

The elements of the $R$-matrix in Equation {\bf \ref{R-matrix}}
are the weights of the vertices of the rational six-vertex model. 
This is the origin of the connection 
of the two models. One can graphically represent the elements 
of (\ref{R-matrix}) to obtain the six vertices of the
rational six-vertex model in Figure {\bf \ref{six-vertices}}. 
Naturally, they satisfy the same properties, namely unitarity, 
crossing symmetry and the crucial Yang-Baxter equations that 
are required for integrability. 
 
\subsection{The $L$-matrix}

The $L$-matrix of the XXX spin chain is a local operator 
that acts non-trivially on one site of the spin chain only. 
It acts non-trivially on the auxiliary space $h_a$ and on 
the $i$-th quantum space, and acts trivially all other quantum 
spaces. 
The mechanics of the construction and the precise action of 
the $L$-matrix require more space than we can afford in this 
note. We refer the reader to \cite{korepin-book} for a detailed 
exposition.

\subsection{The Monodromy matrix}

The monodromy matrix is a global operator that acts on all 
sites in the spin chain. It is constructed as an ordered 
direct product of the $L$-matrices that act on single sites. 
It is typically written in $(2 \! \times  \! 2)$ block form as 

\begin{align}
M_{a}(x, \{z\}_L)
=
\ll
\begin{array}{cc}
A(x) & B(x) \\
C(x) & D(x)
\end{array}
\rr_a
\end{align}

\noindent where the matrix entries are operators that act in 
${\mathcal{H}} = h_1 \otimes \cdots \otimes h_L$. To simplify 
the notation, we have omitted the dependence of the elements 
of the $M$-matrix on the quantum rapidities $\{z\}$. This 
dependence is implied from now on. For the purposes of this 
note, the main aspect of the elements of the $M$-matrix 
that we need to know is that they can represented in 
six-vertex model terms as the horizontal lines in Figure 
{\bf \ref{four-lines}}.
The $A$, $B$, $C$ and $D$-lines are the six-vertex model 
representation of the corresponding elements of the $M$-matrix. 
This representation is very useful and that is why we in 
introduced it in Section {\bf \ref{section-six-vertex}}.

\subsection{Initial and final generic Bethe states}
An initial (final) generic Bethe state is represented in six-vertex 
model terms as a $B$-configuration ($C$-configura\-tion), as defined 
in Section {\bf \ref{section-six-vertex}} and illustrated on 
left (right) hand side of Figure {\bf \ref{initial-final-state}}. 
Note that the outcome of the action of the $N$ $B$-lines 
($C$-lines)
on the initial (final) length-$L$ reference state produces 
a final (initial) spin system that can assume all possible 
spin states of net spin $(L-N)$. Each of these definite spin 
states is weighted by the weight of the corresponding lattice 
configuration (where when sums over all spins on the bulk segments).

\subsection{Bethe eigenstates and Bethe equations}
The initial and final reference states 
$|       [ L^{\wedge} ] \>$ and 
$\< [ L^{\wedge} ] |$ are 
eigenstates of the diagonal elements of the monodromy matrix.
The eigenvalues are easy to compute in terms of the vertex weights 
and will not be listed here as we will not need them. We refer the 
reader to \cite{KMT,MW} for these details. 
This makes these states eigenstates of the transfer matrix $T(x)$, 
which by definition is the trace of the monodromy $M$-matrix, 
that is $T(x)$ $=$ $\Tr \ll M(x) \rr$. 
The rest of the eigenstates $\{ \O \}$ of $T(x)$, that is 

\begin{equation}
    T(x)                |\O \>_{\beta}
=
\ll A(x) + D(x) \rr     |\O \>_{\beta} 
= 
E_{\O}(x)               |\O \>_{\beta}
\label{eigenstate}
\end{equation}

\noindent where $E_{\O}(x)$ is the corresponding eigenvalue, are 
generated using the BA,
which is the statement that all eigenstates 
of $T(x)$ are created in two steps. \1 One acts on the initial 
reference state with the $B$-element of the monodromy 
matrix

\begin{equation}
|\O \>_{\beta} = 
B(u_{\beta N}) \cdots B(u_{\beta 1}) | [ L^{\wedge} ] \>
\end{equation}

\noindent where $N \leq L$, since acting on $| [L^{\wedge} ] \>$ 
with more $B$-operators than the number of sites in the spin chain 
annihilates it.  
\2 We require that the auxiliary space rapidity variables  
$\{
u_{\beta 1}, \cdots, 
u_{\beta N}
\}$ satisfy Bethe equations, hence the use of the subscript $\beta$. 
That is, $|\O \>$ as well as $\< \O|$ are eigenstates of 
$T(x)$ if and only if 

\begin{equation}
\label{bethe-equations}
\prod_{j=1}^{L}
\frac{
a[u_i, z_j]
}{
b[u_i, z_j]
}
=
\prod_{j \not= i}^{N}
\frac{
b[u_j, u_i]
}{
b[u_i, u_j]
}
\end{equation}
\noindent for all $1\leq i \leq N$. Eigenstates of the transfer 
matrix $T(x)$ are also eigenstates of the spin-chain Hamiltonian 
\cite{korepin-book}. The latter is the spin-chain version of the 
1-loop dilatation operator in SYM$_4$. We construct 
eigenstates of $T(x)$ in $\mathcal{H}^{*}$ using the $C$-element 
of the $M$-matrix 

\begin{equation}
_{\beta}\< \O|
=
\< [ L^{\wedge}] |
C(u_{\beta 1})
\ldots
C(u_{\beta N})
\end{equation}
\noindent where $N \leq L$ to obtain a non vanishing result, 
and requiring that the auxiliary space rapidity variables 
satisfy the Bethe equations.

\subsection{A sequence of scalar products that can be evaluated 
as determinants}
Following \cite{KMT,MW}, we define the scalar product 
$S[L, N_1, N_2]$, 
$0 \leq N_2 \leq N_1$, that involves $(N_1 + N_2)$ operators, 
$N_1$ $B$-operators with auxiliary rapidities that satisfy Bethe 
equations, and 
$N_2$ $C$-operators with auxiliary rapidities that are 
free\footnote{To avoid a proliferation of notation, we use $N_1$, 
$N_2$ and $N_3 = N_1 - N_2$, instead of the corresponding
notation used in \cite{KMT,MW}. The reason is that 
these variables will match the corresponding ones in Section 
{\bf \ref{section-structure-constants}}.}. 
For $N_2 = 0$, we obtain, up to a non-dynamical factor, the domain 
wall partition function. For $N_2 = N_1$, we obtain Slavnov's scalar 
product. These scalar products $S[L, N_1, N_2]$ can be found in 
\cite{KMT,MW}
The purpose of the exercise is to show that $S[L, N_1, N_2]$ is 
the partition function (weighted sum over all 
internal configurations) of the $[L, N_1, N_2]$-configurations 
introduced in Section {\bf{\ref{section-six-vertex}}}.

Let 
$\{u\}_{\beta N_1}$ $=$ $\{u_{\beta 1}, \cdots, u_{\beta N_1}\}$, 
$\{v\}_{N_2}$ $=$ $\{v_1, \cdots, v_{N_2}\}$, 
$\{z\}_L$ $=$ $\{z_1, \cdots, z_L\}$ 
be three sets of variables the first of which satisfies Bethe equations, 
$0 \leq N_2 \leq N_1$ and $1\leq N_1 \leq L$. 
We wish to define the scalar products 

\begin{multline}
\label{restricted-scalar-product}
S[L, N_1, N_2] 
\ll 
\{u\}_{\beta N_1}, 
\{v\}_{N_2},
\{z\}_{L  }
\rr
=
\\
\< [ N_3^{\vee}, (L-N_3)^{\wedge}] |
\prod_{i=1}^{N_2} C(v_i) \prod_{j=1}^{N_1} B(u_{\beta j}) 
|[ L^{\wedge} ] \>
\end{multline}

\noindent where $0 \leq N_2 \leq N_1$, 
$N_3 = N_1 - N_2$ 
\footnote{Our choice of vertex weights in 
Equation ({\bf \ref{weights}}), is such that our $B$ and $C$ 
operators as in Equation ({\bf \ref{restricted-scalar-product}})
are the same as the normalized $\mathbb{B}$ and $\mathbb{C}$ 
operators of \cite{KMT}. 
Our expression for the restricted Slavnov product 
in Equation ({\bf \ref{restricted-scalar-product}}) 
agrees with that in \cite{KMT}.}. It is clear that for $N_2=0$, 
we obtain a domain wall partition function, while for $N_2=N_1$, 
we obtain Slavnov's scalar product. 
In all cases, we assume that the auxiliary rapidities  
$\{u\}_{\beta N_1}$ obey the Bethe equations 
(\ref{bethe-equations}), and use the subscript $\beta$ 
to emphasize that, while the auxiliary rapidities $\{v\}_{N_2}$ 
are either free or also satisfy their own set of Bethe equations. 
When the latter is the case, this fact is not used. The quantum 
rapidities $\{z\}_L$ do not satisfy Bethe equations, and are 
taken to be equal to the same constant value in the homogeneous 
limit. 

\subsection{The scalar products $S[L, N_1, N_2]$ are 
$[L, N_1, N_2]$-configurations}
From the definition of $S[L, N_1, N_2]$, one can easily 
identify them as the BA versions of the six-vertex 
$[L, N_1, N_2]$-configurations. We will use this 
fact from now on.

\subsection{A determinant expression for the 
$[L, N_1, N_2]$-restricted Slavnov scalar product}
Following \cite{KMT,MW}, we consider the $(N_1 \! \times \! N_1)$ 
matrix 

\begin{multline}
\mathcal{S}
\ll 
\{u\}_{\beta N_1},
\{v\}_{N_2},
\{z\}_L 
\rr
=
\\
\ll   
\begin{array}{cccccc}
f_{  1}(z_1) & \cdots & f_{  1}(z_{N_3}) & g_{  1}(v_1) & \cdots & g_{  1}(v_{N_2})
\\
\vdots       &        & \vdots           & \vdots       &        & \vdots
\\
f_{N_1}(z_1) & \cdots & f_{N_1}(z_{N_3}) 
                                         & g_{N_1}(v_1) 
				                        & \cdots & g_{N_1}(v_{N_2})
\end{array}
\rr
\end{multline}

\begin{multline}
\nonumber
\! \! \! \! \! \! f_i(z_j)=
\ll \frac{ \eta }{(u_i - z_j + \eta) (u_i - z_j)} \rr
\prod_{k=1}^{N_2}
\frac{1}{(v_k - z_j)} 
\\
g_i(v_j)=
\ll
\frac{\eta}{u_i - v_j}
\rr
\ll   
\ll
\prod_{k=1}^{L} 
\frac{(v_j - z_k + \eta)}{(v_j - z_k)}
\prod_{k \not = i}^{N_1} (u_k - v_j + \eta)  
\rr
-
\prod_{k \not = i}^{N_1} (u_k - v_j - \eta) 
\rr
\end{multline}

\noindent where $N_3 = N_1 - N_2$. 
Since the auxiliary rapidities $\{u\}_{\beta N_1}$ satisfy 
Bethe equations (\ref{bethe-equations}), following 
\cite{KMT,MW} 

\begin{equation}
\label{restricted-slavnov}
S[ L, N_1, N_2] =
\frac{\N_S}{\D_S}
\det \mathcal{S} \ll \{u\}_{\beta N_1}, \{v\}_{N_2}, \{z\}_L \rr 
\end{equation}

\begin{multline}
\N_S = \prod_{i=1}^{N_1} \prod_{j=1}^{N_3} (u_i - z_j + \eta), 
\\
\D_S = 
\prod_{1 \leq i < j \leq  N_1} (u_j-u_i)
\prod_{1 \leq i < j \leq  N_2} (v_i-v_j) 
\prod_{1 \leq i < j \leq  N_3} (z_i-z_j)
\end{multline}

\noindent To conclude, we have a determinant expression for the 
$[L, N_1, N_2]$-configurations introduced in Section 
{\bf{\ref{section-six-vertex}}}. For the result in this note, 
we need the homogeneous limit of $S^{\textit{hom}}[L, N_1, N_2]$. 
Taking the limit $z_i \rightarrow z$, $i \in \{1, \cdots, L\}$, 
the result is

\begin{equation}
S^{\textit{hom}}[L, N_1, N_2] 
=
\frac{
\prod_{i=1}^{N_1} (u_i - z + \eta)^{N_3}
\det
\mathcal{S}^{\textit{hom}} 
\ll \{u\}_{\beta N_1}, \{v\}_{N_2}, z \rr
}
{
\prod_{1 \leq i < j \leq  N_1 } (u_j - u_i)
\prod_{1 \leq i < j \leq  N_2 } (v_i - v_j)
}
\label{restricted-slavnov-homogeneous}
\end{equation}

\begin{multline}
\nonumber
\mathcal{S}^{\textit{hom}} 
\ll 
\{u\}_{\beta N_1},
\{v\}_{N_2}, 
  z 
\rr
=
\\
\ll   
\begin{array}{cccccc}
\Phi^{(0)}_{  1}(z) & \cdots & \Phi^{(N_3 - 1)}_{ 1}(z) & 
g^{\textit{hom}}_{  1}(v_{N_2}) & \cdots & g^{\textit{hom}}_{ 1}(v_1)
\\
\vdots & & \vdots & \vdots & & \vdots
\\
\Phi^{(0)}_{N_1}(z) & \cdots & \Phi^{(N_3 - 1)}_{N_1}(z) & 
g^{\textit{hom}}_{N_1}(v_{N_2}) & \cdots & g^{\textit{hom}}_{N_1}(v_1)
\end{array}
\rr
\end{multline}

\begin{multline}
\nonumber
\Phi^{(j)}_{i} = \frac{1}{j!}
\ \ 
\partial^{(j)}_z
\ \ 
f_i (z),
\\
g^{\textit{hom}}_i(v_j)=
\frac{\eta}{(u_i - v_j)}
\ll
\ll
\frac{
v_j - z + \eta
}
{
v_j - z
}
\rr^L
\prod_{k \not = i}^{N_1} (u_k - v_j   + \eta)
-
\prod_{k \not = i}^{N_1} (u_k - v_j   - \eta)
\rr
\end{multline}

\subsection{The Gaudin norm}
Let us consider the original, unrestricted Slavnov scalar 
product, 
$S[L, N_1, N_2 = N_1, N_3 = 0] 
\ll \{u\}_{\beta N_1}, \{v\}_{N_1}, \{z\}_L \rr$, 
and set $\{v\}_{N_1}$ $=$ $\{u\}_{\beta N_1}$ to obtain the 
Gaudin norm 
$N(\{u\}_{\beta N_1})$ which is the square of the norm of 
the Bethe eigenstate with auxiliary rapidities $\{u\}_{\beta N_1}$. 
It inherits a determinant expression that can be computed 
starting from that of the Slavnov scalar product that we begin 
with and taking the limit $\{v\}_{N_1} \rightarrow \{u\}_{\beta N_1}$.
Using $N_i$ for $N_1$, and following \cite{KMT}, one obtains 

\begin{multline}
\N[L_i, N_i]
\ll
\{u\}_{\beta N_i}, \{z\}_{L_i}
\rr 
= 
{\eta}^{N_i}
\ll
\product_{\alpha \neq \beta}
\frac{u_i - u_j + \eta}
     {u_i - u_j       }
\rr
\det \Phi^{\prime} \ll \{u\}_{\beta N} \rr
\\
\Phi^{\prime}_{ij} \ll \{u\}_{\beta N} \rr
=
- 
\partial_{u_j}
\ln 
\ll 
\ll 
\frac{u_i + z}{u_i - z}
\rr^{L}
\product_{ \begin{array}{c} k = 1 \\ k \neq i \end{array}}^{N}
\frac{u_k - u_i + \eta}
     {u_k - u_i - \eta}
\rr
\label{Gaudin}
\end{multline}

\noindent We need the Gaudin norm to normalize the Bethe 
eigenstates that form the 3-point functions whose structure 
constants we are interested in. 

In Section {\bf \ref{section-six-vertex}}, we learned how to 
construct six-vertex model configurations, using horizontal 
lines that effectively act on vertical line segments with spin 
assignments, and defined the $[L, N_1, N_2]$-configurations. 
In this section, we saw that all objects introduced in 
Section {\bf \ref{section-six-vertex}} have spin-chain 
analogues, and that the scalar products $S[L, N_1, N_2]$ 
are partition functions of the $[L, N_1, N_2]$-configurations,
and that they can be evaluated in determinant form.
In the following section, we will see that the structure 
constants $c^{(0)}_{ijk}$ are nothing but $S[L, N_1, N_2]$ 
scalar products, up to simple factors.

\section{The structure constants of SYM$_4$}
\label{section-structure-constants}

In this section, we discuss the EGSV expression for the structure 
constants in view of what learned in Sections 
{\bf \ref{section-six-vertex}} and 
{\bf \ref{section-spin-chain}}. 

%
\begin{figure}
\thicklines
\setlength{\unitlength}{0.0008cm}
\begin{picture}(14000,6000)(-1000,-1000)
%
\drawline[-30](2400,0300)(4800,2700)
%
%
\path(1200,0300)(3600,2700)
\path(1300,0300)(3700,2700)
\path(1400,0300)(3800,2700)
\path(1500,0300)(3900,2700)
\path(1600,0300)(4000,2700)
\path(1700,0300)(4100,2700)
\path(1800,0300)(4200,2700)
\path(1900,0300)(4300,2700)
\path(2000,0300)(4400,2700)
\path(2100,0300)(4500,2700)
\path(2200,0300)(4600,2700)
\path(3600,2700)(3600,3900)
\path(3700,2700)(3700,3900)
\path(3800,2700)(3800,3900)
\path(3900,2700)(3900,3900)
\path(4000,2700)(4000,3900)
\path(4100,2700)(4100,3900)
\path(4200,2700)(4200,3900)
\path(4300,2700)(4300,3900)
\path(4400,2700)(4400,3900)
\path(4500,2700)(4500,3900)
\path(4600,2700)(4600,3900)
%
%
\path(5600,2700)(8000,0300)
\path(5800,2700)(8200,0300)
\path(6400,2700)(8800,0300)
\path(6600,2700)(9000,0300)
\path(7200,2700)(9600,0300)
\path(5600,2700)(5600,3900)
\path(5800,2700)(5800,3900)
\path(6400,2700)(6400,3900)
\path(6600,2700)(6600,3900)
\path(7200,2700)(7200,3900)
\drawline[-30](4800,2700)(7200,0300)
\drawline[-40](4800,3900)(4800,2700)
%
\path(3600,3900)(3600,2700)(1200,300)
\path(7200,3900)(7200,2700)(9600,300)
%
\path(4800,1500)(3600,0300)
\path(4800,1500)(6000,0300)
%
\path(6000,0300)(9600,0300)
%
\path(1200,0300)(3600,0300)
\path(3600,3900)(7200,3900)
\put( 2900, 4000){$l_1$}
\put( 2900, 2700){$m_1$}
\put( 0500,-0100){$l_3$}
\put( 7400, 4000){$r_1$}
\put( 7400, 2700){$m_3$}
\put( 9800,-0100){$r_2$}
\put( 4900, 4000){$c_1$}
\put( 4900, 2700){$m_0$}
\put( 2400,-0100){$c_3$}
\put( 7200,-0100){$c_2$}
\put( 3600,-0100){$r_3$}
\put( 4800, 0800){$m_2$}
\put( 6000,-0100){$l_2$}
\end{picture}
%
\caption{A schematic representation of a 3-point function. 
State $\O_1$ is at the top. 
State $\O_2$ is at the bottom to the right. 
State $\O_3$ is at the bottom to the left. 
For further details, please see the text.}
%
\label{pants}
\end{figure}
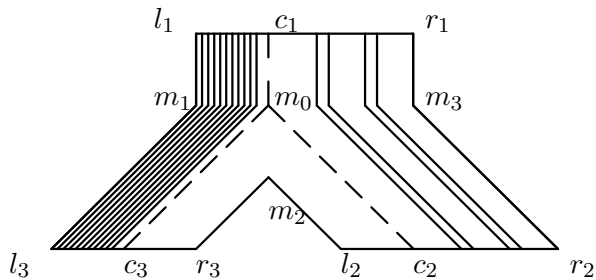
\bigskip

\subsection{Single-trace operators, normalization factors 
and pants diagrams}
Following \cite{E1}, we consider gauge-invariant local
single-trace operators $\{\O\}$, with 1-loop conformal 
dimensions $\{\Delta_{\O}\}$, that consist of two charged 
scalar fields that are not conjugates, and thereby map to 
Bethe eigenstates of an XXX spin-$\frac{1}{2}$ chain. 
For example, a single-trace operators in the $SU(2)$ 
sector spanned by the charged scalars $\{Z, X\}$, 
is in the form $\Tr(ZZXZZZXXZ \cdots)$.

Any 2-point function of two   operators in $\{\O\}$ is 
in the form in Equation ({\bf \ref{2-point}}). 
Any 3-point function of three operators in $\{\O\}$ is 
in the form in Equation ({\bf \ref{3-point}}).  
We choose the normalization factor $\N_i$ to be the Gaudin 
norm of the corresponding Bethe eigenstate

\begin{equation}
\N_i = \N [N_i, L_i] \ll \{u\}_{\beta N_i}, \{z\}_{L_i} \rr
\label{normalization-factor}
\end{equation}

We represent the 3-point functions that we consider in terms 
of a `pants diagram'. Consider the schematic diagram in Figure 
{\bf \ref{pants}}. Identify the pairs of corner points
$\{l_1, r_1\}$,
$\{l_2, r_2\}$,
$\{l_3, r_3\}$,
as well as the triple 
$\{m_1, m_2, m_3\}$ to obtain a pants diagram.

\subsection{Perturbative expansion of structure constants}
The structure constants of these operators have a perturbative 
expansion in the \lq t Hooft coupling constant $\lambda$, 

\begin{equation}
C_{ijk} =
c_{ijk}^{(0)} + \lambda c_{ijk}^{(1)} + \dots
\end{equation}

We restrict the discussion to the leading coefficient $c_{ijk}^{(0)}$. 
In the limit $\lambda \rightarrow 0$, 
many single-trace operators have the same conformal dimension. This 
degeneracy is lifted at 1-loop level and certain linear combinations 
of single-trace operators have definite 1-loop anomalous conformal 
dimension. 
Remarkably, these linear combinations correspond to eigenstates of 
a closed XXX spin-$\frac{1}{2}$ chain. Their anomalous conformal 
dimensions are the corresponding Bethe eigenvalues. These closed spin 
chain states correspond to the circles at the boundaries of the 
pants diagram that can be constructed from 
Figure {\bf \ref{pants}} as discussed above.

To construct three-point functions at the SYM$_4$ operator level, 
the fundamental scalar fields in the operators 
$\O_i$, $i = \{1, 2, 3\}$ are contracted by free propagators. 
Each propagator connects two fields, hence $L_1 + L_2 + L_3$ 
is an even number. The number of propagators between 
$\O_i$ and $\O_j$ is 
\begin{equation}
l_{ij}= \frac{1}{2} (L_i + L_j - L_k)
\end{equation}
\noindent where $(i,j,k)$ take distinct values in $(1,2,3)$. 
We restrict our attention to the non-extremal case, that is, 
all $l_{ij}$'s are strictly positive. 
Following \cite{E1}, the free propagators reproduce the factor 
$1/|x_i - x_j|^{\Delta_i+\Delta_j-\Delta_k}$ in 
Equation {\bf (\ref{3-point})}, where $\Delta_i$ $=$ $\Delta_i^{(0)}$, 
the tree-level conformal dimension. See Figure {\bf \ref{pants}} 
for a schematic representation of a three point function of the 
type discussed in this note. The horizontal line segment between 
$l_i$ and $r_i$ represents the operator $\O_i$. 
The lines that start at $O_1$ and end at either $\O_2$ or $\O_3$ 
represent one type of propagators. More details are given below.

\subsection{From single-trace operators to spin-chain states} 
One represents the single-trace operator $\O_i$ of well-defined 
1-loop anomalous conformal dimension $\Delta_i$ by a closed 
spin-chain Bethe eigenstate 
$| \O_i \>_{\beta}$. Its eigenvalue $E_i$ is equal to $\Delta_i$. 
The number of fundamental 
fields $L_i$ in the trace is the length of the spin chain. 

The single-trace operator $\O_i$ is a composite operator built
from weighted sums over traces of products of two complex scalar 
fundamental fields $\{X, Z\}$ and their conjugates. These 
fundamental fields are mapped to definite (up and down) spin 
states. A crucial step in \cite{E1} is the identification of 
the operator content of $\O_i$, $i \in \{1, 2, 3\}$ with 
spin-chain spin states as follows. 

\setlength{\extrarowheight}{6.0pt}
\begin{center}
\begin{table}[h]
\begin{tabular}{| c || c | c | c | c |}
\hline
{Operator} & 
{$\ll \begin{array}{c} 1 \\  0 \end{array} \rr_i$} & 
{$\ll \begin{array}{c} 0 \\  1 \end{array} \rr_i$} &
{$\ll                  1 \ \ 0             \rr_i$} &
{$\ll                  0 \ \ 1             \rr_i$} 
\\
\hline
\hline
$\O_1$ & $      Z  $ & $      X  $ & $ \bar{Z} $ & $ \bar{X} $ 
\\ 
\hline
$\O_2$ & $ \bar{Z} $ & $ \bar{X} $ & $      Z  $ & $      X  $ 
\\
\hline
$\O_3$ & $      Z  $ & $ \bar{X} $ & $ \bar{Z} $ & $      X  $ 
\\
\hline
\end{tabular}
\bigskip
\caption{Identification of operator content of $\O_i$,
$i \in \{1, 2, 3\}$ with spin states in initial and final 
spin chain states}
\end{table}
\end{center}

From Table {\bf 1}, one can read the fundamental-scalar 
operator content of each single-trace operator $\O_i$,
$i \in \{1, 2, 3\}$, when it is an initial state and 
when it is a final state.
For example, the fundamental scalar operator content 
of the initial state 
$ |\O_1 \>$ is $\{      Z,       X\}$, 
and that of the corresponding final state 
$\<\O_1  |$ is $\{ \bar{Z}, \bar{X}\}$.
The content of an initial state and the corresponding 
final state are related by the `flipping' operation
described below.

\subsection{Remarks}
\1 Following \cite{E1}, since we can Wick contract 
a scalar $f$ only with its conjugate $\bar{f}$, the above 
is the only choice that is fully contained in the $SU(2)$ 
sector of the theory and involves non-extremal correlators 
at the same time. 
\2 In computing structure constants, we identify the fundamental 
scalar fields with definite spin states only after we write
the structure constants in terms of three scalar products 
and ignore one of them as trivial. It is only then that 
the identification becomes unique and simple.
\subsection{Structure constants in terms of spin chain}
Having mapped the single-trace operators $O_i$, 
$i \in \{1, 2, 3\}$ to spin-chain eigenstates, 
EGSV construct the structure constants in three steps.

\subsection*{Step \1 Split the lattice configurations
that correspond to closed spin chain eigenstates into 
two parts}
Consider the open 1-dimensional lattice configuration 
that corresponds to the $i$-th closed spin chain 
eigenstate, $i \in \{1, 2, 3\}$. 
This is schematically represented by a line in Figure 
{\bf \ref{3-point}} that starts at $l_i$ and ends at $r_i$. 
Split that, at point $c_i$ into 
{\it left} and {\it right} sub-lattice configurations 
of lengths 
$L_{i, L} = \frac{1}{2} (L_i + L_j - L_k)$ and 
$L_{i, R} = \frac{1}{2} (L_i + L_k -L_j)$ 
respectively. Note that the lengths of the sub-lattices 
is fully determined by $L_1$, $L_2$ and $L_3$ which are 
fixed\footnote{EGSV interpret the result of this operation 
as two open spin chains. In this note, we prefer to 
interpret it as two open lattice configuration that 
represents the closed spin-chain eigenstates, and 
stay clear of open spin chains. 
This is because the BA operators 
used throughout are those that act on lattice 
configurations that represent closed spin chain 
states. This is a matter of interpretation, and 
the final technical result remains the same.}. 

Following \cite{korepin-book}, we express the single 
lattice configuration of the original closed spin chain 
state as a weighted sum of tensor products of states 
that live in two smaller Hilbert spaces. The latter 
correspond to closed spin chains of lengths $L_{i, L}$ 
and $L_{i, R}$ respectively. That is, 
$|\O_i\> = \sum H_{L, R} |\O_{i}\>_l \otimes |\O_{i}\>_r$.
The factors $H_{L, R}$ were computed in 
\cite{korepin-book} and were needed in \cite{E1}, 
where one of the scalar products is generic and had 
to be expressed as an explicit sum. They will not be 
needed in this work as we use Bethe equations to 
evaluate this very sum as a determinant.

\subsection*{Step 2. Map initial states to corresponding 
final states.} 
In \cite{E1}, EGSV perform the mapping 
$|\O_i\>_l \otimes |\O_i\>_r\ \to 
 |\O_i\>_l \otimes _r\<\O_i|$, 
using the operator $\F$\footnote{EGSV take pains to explain 
how the flipping operation 
is {\it not} the same as conjugation operation familiar 
from Quantum Mechanics textbooks. We refer the reader 
to \cite{E1} for details. Further, we will {\it not} 
follow the notation of \cite{E1} and add an upper arrow 
to distinguish a flipped state from a conjugated one as 
we will not consider any examples of the latter.}.
that acts as follows.
\begin{equation}
\mathcal{F} 
\ll | 
f_1 f_2 \cdots f_{L-1} f_{L}
\> 
\rr 
= 
\< 
\bar{f}_{L} \bar{f}_{L-1} \cdots \bar{f}_2 \bar{f}_1 | 
\label{flipping}
\end{equation}
\noindent In particular,

\begin{equation}
\<     ZZ \cdots Z|     ZZ \cdots Z\> = 
\<\bar{Z} \bar{Z} \cdots \bar{Z} |   
   \bar{Z} \bar{Z} \cdots \bar{Z} \> = 1, \quad
\<\bar{Z} \bar{Z} \cdots \bar{Z} | Z Z \cdots Z \> 
 = 0 
\end{equation}

\noindent More generally

\begin{equation}
\< f_{i_1} f_{i_2} \cdots f_{i_L} | f_{j_1} f_{j_2} \cdots f_{j_L}  \> 
\sim 
\delta{i_1 j_1} \delta{i_2 j_2} \cdots \delta{i_L j_L} 
\end{equation}

\noindent The `flipping' operation in Equation {\bf \ref{flipping}} 
is the origin of the differences in assignments of fundamental scalar 
fields to initial and final operator states in Table {\bf 1}. 
For example, $| \O_1 \>$ has field content $\{Z, X\}$, but 
$\< \O_1 |$ has field content $\{ \bar{Z}, \bar{X}\}$.
This agrees with the fact that in computing $\< \O_i | \O_i \>$, 
free propagators can only connect charge conjugate scalar fields.

\subsection*{Step 3. Compute scalar products}
The final step is to Wick contract pairs 
of initial states 
$|\O_{i  }\>_r$ 
and final states
$|\O_{i+1}\>_l$, where $i \in \{1, 2, 3\}$ and $i + 3 \equiv i$. 
The spin-chain equivalent of that is to compute the scalar 
products
$\,_r\< \O_i| \O_{i+1}\>_l$,
which in six-vertex model terms are $BC$-configurations.
The most general scalar product that we can consider is the 
generic scalar product between two generic Bethe states

\begin{equation}
S_{\textit{generic}} \ll \{u\},\{v\} \rr = 
\<0 | 
\prod_{j=1}^N \mathcal C(v_j) 
\prod_{j=1}^N \mathcal B(u_j) 
| 0 \> \, 
\end{equation}

A computationally tractable evaluation of 
$S_{\textit{generic}}(\{u\},\{v\})$
using the commutation relations of BA operators is known 
\cite{Korepin}. Simpler expressions are obtained when the 
auxiliary rapidities of one (or both) states satisfies 
Bethe equations. The result in this case is a determinant. 
When only one set satisfies Bethe equations, one obtains
a Slavnov scalar product. This was discussed in Section
{\bf \ref{section-spin-chain}}.

\subsection{A preliminary, unevaluated expression}
The above three steps lead to the following preliminary, 
unevaluated expression

\begin{equation}
c^{(0)}_{123} = 
\N_{123}
\sum \limits_{a,b,c}                            \ 
_r\< {\O}_{3_c}| \O_{1_a}\>_l \ 
_r\< {\O}_{1_a}| \O_{2_b}\>_l \ 
_r\< {\O}_{2_b}| \O_{3_c}\>_l \
\label{intermediate}
\end{equation}

\noindent where the normalization factor, that will turn out 
to be a non-trivial object that depends on the norms of the 
Bethe eigenstates, is 

\begin{equation}
\N_{123} = \sqrt{ \frac{L_1 L_2 L_3} {\N_1 \N_2 \N_3} }
\label{normalization}
\end{equation}

The sum in Equation {\bf \ref{intermediate}} is to be understood
as follows. \1 It is a sum over all possible ways to split 
the sites of each closed spin chain (represented as a segment in 
a 1-dimensional lattice) into a left part and a right part. 
We will see shortly that only one term in this sum survives.
\2 It is a sum over all possible ways of partitioning the $X$ 
or $\bar{X}$ content of a spin chain state between the two parts 
that that spin chain was split into. We will see shortly that 
only one sum will survive.

\subsection{A constraint that leads to simplifications}
Wick contracting single-trace operators, we can only contract 
a fundamental scalar with its conjugate. Given the assignments 
in Table {\bf 1}, one can see that 
\1 {\it All} $Z$ fields in $|\O_3 \>$ contract with $\bar{Z}$ 
fields in $\O_2$. 
The reason is that there are $\bar{Z}$ fields in $\O_2$, 
and none in $\O_1$.
\2 All $\bar{X}$ fields in $\O_3$ contract with $X$ 
fields in $\O_1$. 
The reason is that there $X$ fields only in $\O_1$, 
and none in $\O_2$.
If the total number of scalar fields in $\O_i$ is $L_i$, 
and the number of $\{X, \bar{X}\}$-type scalar fields is 
$N_i$, then 

\begin{equation}
l_{13} = N_3,     \qquad
l_{23} = L_3 - N_3, \qquad 
l_{12} = L_1 - N_3
\label{relations}
\end{equation}

\noindent and, we have the constraint 

\begin{equation}
N_1 = N_2 + N_3 
\label{constraint}
\end{equation}

From 
Equation {\bf \ref{relations}} and 
Equation {\bf \ref{constraint}}, 
we have the following simplifications.
\1 There is only one way to split each lattice 
configuration that represents a spin chain into a left 
part and a right part. 
\2 The scalar product 
$_r\< {\O}_{2_b}| \O_{3_c}\>_l$ 
involves the fundamental scalar field $Z$ (and only $Z$) 
in the initial state $| \O_{3_c}\>_l$ as well as 
in the final   state $_r\< {\O}_{2_b} |$.
Using Table {\bf 1}, 
we find that these states translate to an initial and a 
final reference state, respectively. This is represented 
in Figure {\bf \ref{pants}} by the fact that no connecting 
lines (that stand for propagators of 
$\{X, \bar{X}\}$ states) connect $\O_2$ and $\O_3$.
The scalar product of two reference states is 
${}_r\< {\O}_{2_b}| \O_{3_c}\>_l \ = 1$.

\noindent \3 The scalar product 
$ _r\< {\O}_{1}| \O_{3}\>_l$ 
involves the fundamental scalar fields 
$\bar{X}$ (and only $\bar{X}$)
in the initial state $| \O_{3}\>_l$ as well as
in the final   state $_r\< {\O}_{1}|$.
Using Table {\bf 1},
we find that these states translate to an initial
and a final dual reference state respectively.
This is represented in Figure {\bf \ref{pants}}
by the high density of connecting lines (that stand for
propagators of $\{X, \bar{X}\}$ states) between  
$\O_1$ and $\O_3$.
The scalar product of two dual reference states 
is straightforward to evaluate in terms of domain 
wall partition functions.
In the remaining scalar product 
${}_r \< {\O}_{1}         | \O_{2}\>_l$, 
both the initial state 
$| \O_{2}\>_l$ and the final state ${}_r \< {\O}_{1} |$
involve $\{\bar{X}, \bar{Z} \}$. These states translate 
to up and down spin and the scalar product is generic. 
Using the BA commutation relations, it can be evaluated 
as a weighted sum \cite{korepin-book}.

\subsection{The EGSV expression}
In \cite{E1}, EGSV put the above facts together and 
obtain an expression for $c^{(0)}_{ijk}$ in 
Equation {\bf \ref{intermediate}}, in the form

\begin{equation}
c^{(0)}_{123} = 
\N_{123}
\   
{\F}_1
\   
\sum_{\alpha \cup \bar{\alpha} = \{u\}_{\beta N_1}}
{\F}_2 \ \ 
{}_r \< [ {N_3}^{\vee}]  | \O_{1}\>_l \
{}_r \< {\O}_{1}         | \O_{2}\>_l \
\label{EGSV-expression}
\end{equation}

\noindent where the normalization factor $\N_{123}$ 
is defined in Equation {\bf \ref{normalization}}, 
$_r\< [ {N_3}^{\vee} ]|$ is a dual reference 
state of length $N_3$, and $\F_1$ and $\F_2$ are factors the 
precise form of which need not concern us here\footnote{EGSV 
obtain their expression in a coordinate Bethe Ansatz basis. 
This leads to factors relative to the algebraic Bethe Ansatz 
basis that we use in this note. We collect these factors in 
$\F_1$ and $\F_2$.}.  
The sum in Equation {\bf \ref{EGSV-expression}} is over all 
possible ways to partition the rapidities $\{u\}_{\beta N_1}$ into two 
sets $\alpha$ and $\bar{\alpha}$, with cardinality $N_2$ and 
$N_3$, respectively. In the next section, we organize the 
computation of $c^{(0)}_{ijk}$ differently, and obtain a result 
that evaluates the sum in Equation {\bf \ref{EGSV-expression}}
as a determinant.

\section{A determinant expression for the structure constants}
\label{section-determinant-expression}
The idea of this note is to identify the expression in 
Equation {\bf \ref{intermediate}}, up simple factors, 
with the restricted scalar product $S[L, N_1, N_2]$, which is 
the partition function of an $[L, N_1, N_2]$-configuration, 
and that can be evaluated as a determinant. This requires 
two simple steps.

\subsection{Step 1. Re-writing one of the scalar products}
We use the facts that  
\1 ${}_r\< {\O}_{2} | {\O}_{3} \>_l $ $=$ $1$, and
\2 ${}_r\< {\O}_{2} | {\O}_{1} \>_l $ $=$ 
   ${}_l\< {\O}_{1} | {\O}_{2} \>_r $, 
which is true for all scalar products, to re-write 
Equation {\bf \ref{intermediate}} in the form

\begin{equation}
c^{(0)}_{123} =
\N_{123}
\sum_{\alpha \cup \bar{\alpha} = \{u\}_{\beta N_1}}
{}_r\< \O_3 | \O_1 \>_l \
{}_l\< \O_2 | \O_1 \>_r \
=
\N_{123} \ 
\ll {}_r\< \O_3 | {}_l \bigotimes \< \O_2 | \rr | \O_1 \> \
\label{unevaluated-more-simplified}
\end{equation}

\noindent where the right hand side of 
Equation {\bf \ref{unevaluated-more-simplified}} 
is a scalar product of the full initial state 
$| \O_1 \> $ (so we no longer have a sum over partitions
of the rapidities $\{u\}_{\beta N_1}$ since we no longer 
split the state $\O_1$) and two states that are pieces of 
original states that were split. This right hand 
side is identical to an $[L, N_1, N_2]$-configuration, 
apart from the fact that it includes an 
$(N_3 \times N_3)$-domain wall configuration, 
that corresponds to the dual reference state contribution of 
${}_r\< [ {N_3}^{\vee} ] |$, 
that is not included in an $[L, N_1, N_2]$-configuration. 

\subsection{Step 2. Accounting for the domain wall partition
functions}
Accounting for the domain wall partition function, and working 
in the homogeneous limit where all quantum rapidities are set 
to $z = \frac{1}{2} \sqrt{-1}$, we 
obtain our result for the structure constants, which 
up to a factor, is in determinant form.

\begin{equation}
\boxed{
c^{(0)}_{123} =
\N_{123}
\ \ 
Z_{N}^{\textit{hom}} \ll \{w\}_{N_3}, \frac{1}{2} \sqrt{-1} \rr
\ \ 
S^{\textit{hom}}[L, N_1, N_2] 
\ll \{u\}_{\beta N_1}, \{v\}_{N_2}, \frac{1}{2} \sqrt{-1}\rr
}
\label{evaluated}
\end{equation}
\bigskip

\noindent where the normalization
$\N_{123}$ is defined in Equation {\bf \ref{normalization}},
the $(N_3 \! \times \! N_3)$ domain wall partition function 
$Z_{N}^{\textit{hom}} \ll \{w\}_{N_3}, \frac{1}{2} \sqrt{-1} \rr$ 
is given in Equation {\bf \ref{izergin-homogeneous}}. 
$S^{\textit{hom}}[L, N_1, N_2]$ 
$\ll \{u\}_{\beta N_1}, \{v\}_{N_2}, \frac{1}{2} \sqrt{-1}\rr$,
is an $(N_1 \! \times \! N_1)$ determinant expression of 
the partition function of an $[L, N_1, N_2]$-configuration, 
given in Equation {\bf \ref{restricted-slavnov-homogeneous}}.
Notice that $\{v\}$ and $\{w\}$ are actually 
            $\{v\}_{\beta}$ and $\{w\}_{\beta}$, that is, they 
	    satisfy Bethe equations, but this fact is not used.

The auxiliary rapidities 
$\{u\}$,
$\{v\}$, and 
$\{w\}$, are those of the eigenstates 
$\O_1$,
$\O_2$, and 
$\O_3$, in \cite{E1}, respectively.

%
\begin{figure}
\setlength{\unitlength}{0.0009cm}
\begin{picture}(10000,08000)(000,2000)
\thicklines
%
%
\path(0600,7500)(9600,7500)
\path(0600,6900)(9600,6900)
\path(0600,6300)(9600,6300)
\path(0600,5700)(9600,5700)
\path(0600,5100)(9600,5100)
\drawline[-30](0000,4500)(9600,4500)
\path(0600,3900)(3000,3900)
\path(0600,3300)(3000,3300)
\path(0600,2700)(3000,2700)
\path(3600,3900)(9600,3900)
\path(3600,3300)(9600,3300)
%
%
\path(1200,8100)(1200,2100)
\path(1800,8100)(1800,2100)
\path(2400,8100)(2400,2100)
\drawline[-30](3300,8700)(3300,1500)
\path(4200,8100)(4200,2700)
\path(4800,8100)(4800,2700)
\path(5400,8100)(5400,2700)
\path(6000,8100)(6000,2700)
\path(6600,8100)(6600,2700)
\path(7200,8100)(7200,2700)
\path(7800,8100)(7800,2700)
\path(8400,8100)(8400,2700)
\path(9000,8100)(9000,2700)
\blacken\path(0900,3990)(0600,3900)(0900,3810)(0900,3990)
\blacken\path(0900,3390)(0600,3300)(0900,3210)(0900,3390)
\blacken\path(0900,2790)(0600,2700)(0900,2610)(0900,2790)
\blacken\path(3900,3990)(3600,3900)(3900,3810)(3900,3990)
\blacken\path(3900,3390)(3600,3300)(3900,3210)(3900,3390)
\blacken\path(0900,5010)(1200,5110)(0900,5190)(0900,5010)
\blacken\path(0900,5610)(1200,5700)(0900,5790)(0900,5610)
\blacken\path(0900,6210)(1210,6300)(0900,6390)(0900,6210)
\blacken\path(0900,6810)(1200,6900)(0900,6990)(0900,6810)
\blacken\path(0900,7410)(1200,7500)(0900,7590)(0900,7410)
\blacken\path(1300,7800)(1210,8100)(1120,7800)(1300,7800)
\blacken\path(1900,7800)(1810,8100)(1720,7800)(1900,7800)
\blacken\path(2500,7800)(2410,8100)(2320,7800)(2500,7800)
%
\blacken\path(4300,7800)(4210,8100)(4120,7800)(4300,7800)
\blacken\path(4900,7800)(4810,8100)(4720,7800)(4900,7800)
\blacken\path(5500,7800)(5410,8100)(5320,7800)(5500,7800)
\blacken\path(6100,7800)(6010,8100)(5920,7800)(6100,7800)
\blacken\path(6700,7800)(6610,8100)(6520,7800)(6700,7800)
\blacken\path(7300,7800)(7210,8100)(7120,7800)(7300,7800)
\blacken\path(7900,7800)(7810,8100)(7720,7800)(7900,7800)
\blacken\path(8500,7800)(8410,8100)(8320,7800)(8500,7800)
\blacken\path(9100,7800)(9010,8100)(8920,7800)(9100,7800)
\blacken\path(9300,7600)(9000,7510)(9300,7420)(9300,7600)
\blacken\path(9300,7000)(9000,6910)(9300,6820)(9300,7000)
\blacken\path(9300,6400)(9000,6310)(9300,6220)(9300,6400)
\blacken\path(9300,5800)(9000,5710)(9300,5620)(9300,5800)
\blacken\path(9300,5200)(9000,5110)(9300,5020)(9300,5200)
\blacken\path(9300,3800)(9600,3890)(9300,3980)(9300,3800)
\blacken\path(9300,3200)(9600,3290)(9300,3380)(9300,3200)
%
%
\path(-0360,2700)(0000,2700)
\path(-0360,3300)(0000,3300)
\path(-0360,3900)(0000,3900)
\put(-1000,3900){$w_{N_3}$}
\put(-1000,2700){$w_1$}
\put(9900,3900){$v_{N_2}$}
\put(9900,3300){$v_1$}
\whiten\path(0000,2610)(0360,2700)(0000,2790)(0000,2610)
\whiten\path(0000,3210)(0360,3300)(0000,3390)(0000,3210)
\whiten\path(0000,3810)(0360,3900)(0000,3990)(0000,3810)
\path(-0360,5100)(0000,5100)
\path(-0360,5700)(0000,5700)
\path(-0360,6300)(0000,6300)
\path(-0360,6900)(0000,6900)
\path(-0360,7500)(0000,7500)
\put(-1000,5100){$u_{N_1}$}
\put(-1000,7500){$u_1$}
\whiten\path(0000,5010)(0360,5100)(0000,5190)(0000,5010)
\whiten\path(0000,5610)(0360,5700)(0000,5790)(0000,5610)
\whiten\path(0000,6210)(0360,6300)(0000,6390)(0000,6210)
\whiten\path(0000,6810)(0360,6900)(0000,6990)(0000,6810)
\whiten\path(0000,7410)(0360,7500)(0000,7590)(0000,7410)
%
%
\path(1200,9220)(1200,8500)
\path(1800,9220)(1800,8500)
\path(2400,9220)(2400,8500)
%
\path(4200,9220)(4200,8500)
\path(4800,9220)(4800,8500)
\path(5400,9220)(5400,8500)
\path(6000,9220)(6000,8500)
\path(6600,9220)(6600,8500)
\path(7200,9220)(7200,8500)
\path(7800,9220)(7800,8500)
\path(8400,9220)(8400,8500)
\path(9000,9220)(9000,8500)
\put(1100,9500){$z_{1}$}
\put(2300,9500){$z_{N_3}$}
\put(8900,9500){$z_{L}$}
\whiten\path(1290,8860)(1200,8500)(1110,8860)(1290,8860)
\whiten\path(1890,8860)(1800,8500)(1710,8860)(1890,8860)
\whiten\path(2490,8860)(2400,8500)(2310,8860)(2490,8860)
%
\whiten\path(4290,8860)(4200,8500)(4110,8860)(4290,8860)
\whiten\path(4890,8860)(4800,8500)(4710,8860)(4890,8860)
\whiten\path(5490,8860)(5400,8500)(5310,8860)(5490,8860)
\whiten\path(6090,8860)(6000,8500)(5910,8860)(6090,8860)
\whiten\path(6690,8860)(6600,8500)(6510,8860)(6690,8860)
\whiten\path(7290,8860)(7200,8500)(7110,8860)(7290,8860)
\whiten\path(7890,8860)(7800,8500)(7710,8860)(7890,8860)
\whiten\path(8490,8860)(8400,8500)(8310,8860)(8490,8860)
\whiten\path(9090,8860)(9000,8500)(8910,8860)(9090,8860)
%
%
%
%
%
%
\blacken\path(2640,3810)(3000,3900)(2640,3990)(2640,3810)
\blacken\path(2640,3210)(3000,3300)(2640,3390)(2640,3210)
\blacken\path(2640,2610)(3000,2700)(2640,2790)(2640,2610)
%
%
\blacken\path(1300,2340)(1210,2700)(1120,2340)(1300,2340)
\blacken\path(1900,2340)(1810,2700)(1720,2340)(1900,2340)
\blacken\path(2500,2340)(2410,2700)(2320,2340)(2500,2340)
%
%
%
%
%
\blacken\path(1300,4860)(1210,4500)(1120,4860)(1300,4860)
\blacken\path(1900,4860)(1810,4500)(1720,4860)(1900,4860)
\blacken\path(2500,4860)(2410,4500)(2320,4860)(2500,4860)
%
%
\blacken\path(1300,4260)(1210,3900)(1120,4260)(1300,4260)
\blacken\path(1900,4260)(1810,3900)(1720,4260)(1900,4260)
\blacken\path(2500,4260)(2410,3900)(2320,4260)(2500,4260)
%
\blacken\path(4300,2940)(4210,3300)(4120,2940)(4300,2940)
\blacken\path(4900,2940)(4810,3300)(4720,2940)(4900,2940)
\blacken\path(5500,2940)(5410,3300)(5320,2940)(5500,2940)
\blacken\path(6100,2940)(6010,3300)(5920,2940)(6100,2940)
\blacken\path(6700,2940)(6610,3300)(6520,2940)(6700,2940)
\blacken\path(7300,2940)(7210,3300)(7120,2940)(7300,2940)
\blacken\path(7900,2940)(7810,3300)(7720,2940)(7900,2940)
\blacken\path(8500,2940)(8410,3300)(8320,2940)(8500,2940)
\blacken\path(9100,2940)(9010,3300)(8920,2940)(9100,2940)
\end{picture}
%
\caption{The six-vertex lattice configuration that corresponds, up to 
a normalization factor $\N_{123}$, to the structure constant 
$c^{(0)}_{123}$.} 
%
\label{lattice-3-pt}
\end{figure}
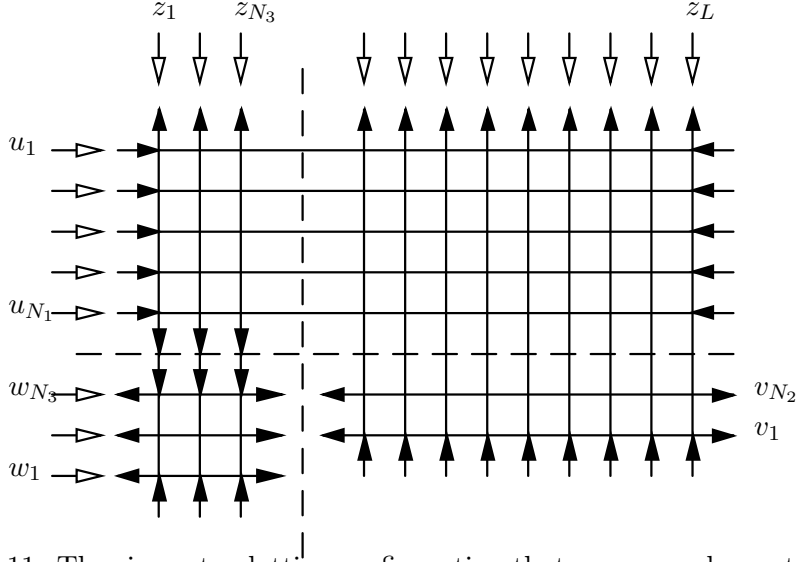
\bigskip

\section{Comments}
\label{section-comments}

Let us consider Figure {\bf \ref{lattice-3-pt}} which 
shows the six-vertex representation of $c^{(0)}_{123}$, 
after a trivial scalar product between two reference 
states (one came from part of state $O_2$ and the other 
from part of state $O_3$) is ignored.
In \cite{E1}, EGSV split all three states, so they split 
state $\O_1$ as well. This splitting is represented by the
vertical dashed line in Figure {\bf \ref{lattice-3-pt}}.
Next they proceed to evaluate the two scalar products
(the third is trivial). The $C$ operators in the two 
final (partial) states are well segregated. But the 
$N$ $B$ operators of the initial state $\O_1$ must 
be partitioned into two sets. 
One of cardinality $(N_3 = N_1-N_2)$ to match the $C$ 
operators from the remainder of $\O_2$, and one of 
cardinality $N_2$ to match the $C$ operators from the 
remainder of $\O_3$. 
There is no unique way to do this, and one can show 
explicitly that one has to sum over all partitions
of the auxiliary rapidities $\{u\}$ of $\O_1$. This 
is the origin of the sum in EGSV expression.

In this note, we do not split $\O_1$, but we identify 
the configuration in Figure {\bf \ref{lattice-3-pt}}
as (up to minor modifications) an object that has 
a known partition function that can be expressed as 
a determinant.
Another way to say it is that by not splitting $\O_1$, 
it remained a Bethe eigenstate and we have effectively 
used the Bethe equations to put the partition function
in determinant form. The Bethe equations play 
a crucial role in the proof of the determinant form 
of this partition function \cite{KMT,MW}.

In \cite{E2}, the limit where one of the operators 
is much smaller than the other two was considered. 
A precise match between weak and strong coupling in 
the Frolov-Tseytlin classical limit for a general 
class of classical solutions was obtained. 
In \cite{E3}, 3-point functions between one large 
classical operator and two large BPS operators were 
computed at weak coupling. 
In \cite{Jan}, a multiple integral expression for 
the generic scalar product, and from that a multiple 
integral version of the EGSV expression was obtained. 
In \cite{plefka}, a systematic perturbative study 
of 3-point functions at 1-loop level, involving 
single-trace operators up to length five, was 
performed.  
In \cite{Pedro}, a non-trivial numerical check showed 
that the result in this note agrees with the EGSV 
expression in \cite{E1}.

\section*{Acknowledgments}
\label{section-acknowledgements}

I wish to thank I Aniceto for introducing me to the topic of this note, 
N Gromov, P Vieira and M Wheeler for patiently explaining their work 
to me, C Ahn, S McAteer, R Nepomechi and M Wheeler for comments 
that helped me improve the manuscript and in coding the determinant 
expression to compare it numerically with the EGSV sum expression, 
and P Vieira for confirming that the two expressions agree. I also 
wish to thank CERN, the Perimeter Institute and the Kavli Institute 
for Theoretical Physics for stimulating research environments, and 
the Australian Research Council for financial support.

\end{document}